\documentclass[11 pt]{article}
\usepackage[utf8]{inputenc}
\usepackage{amsmath}
\usepackage{amssymb}
\usepackage{multirow}
\usepackage[margin=1in]{geometry}
\usepackage{graphicx}
\usepackage{caption}
\usepackage{subcaption}
\usepackage{xcolor}
\usepackage{verbatim}
\usepackage{bigints}

\pdfoutput=1
\usepackage{jheppub}
\usepackage{amsthm}
\usepackage{hyperref}

\usepackage{physics}
\usepackage{xcolor}
\usepackage{mathtools}
\usepackage{tensor}
\usepackage{empheq}

\usepackage{tikz}
\usetikzlibrary{calc,fadings,decorations.pathreplacing}

\usepackage{url}

\def\be{\begin{equation}}
	\def\ee{\end{equation}}
\def\bea{\begin{eqnarray}}
	\def\eea{\end{eqnarray}}

\usepackage{dsfont}
\def\CC{{\mathds{C}}}
\def\RR{{\mathds{R}}}
\def\ZZ{{\mathds{Z}}}

\newcommand{\fft}[2]{\frac{#1}{#2}}

\preprint{LCTP-22-17}

\title{Entanglement and Topology in RG Flows Across Dimensions: Caps, Bridges and Corners}

\author[a]{Evan Deddo,}

\author[a,b,c]{Leopoldo A.~Pando Zayas,}

\author[d]{Christoph F.~Uhlemann}

\emailAdd{evdedd@umich.edu, lpandoz@umich.edu, uhlemann@maths.ox.ac.uk}

\affiliation[a]{Leinweber Center for Theoretical Physics, 
University of Michigan, Ann Arbor, MI 48109, USA}
\affiliation[b]{School of Natural Sciences, Institute for Advanced Study, Princeton, NJ 08540, USA}

\affiliation[c]{The Abdus Salam International Centre for Theoretical Physics, 34014 Trieste, Italy}
\affiliation[d]{Mathematical Institute, University of Oxford, Oxford  OX2 6GG, United Kindom}

\abstract{We quantitatively address the following question: for a QFT which is partially compactified, so as to realize an RG flow from a $D$-dimensional CFT in the UV to a $d$-dimensional CFT in the IR, how does the entanglement entropy of a small spherical region probing the UV physics evolve as the size of the region grows to increasingly probe IR physics? This entails a generalization of spherical regions to setups without full Lorentz symmetry, and we study the associated entanglement entropies holographically. We find a tight interplay between the topology and geometry of the compact space and the evolution of the entanglement entropy, with universal transitions from `cap' through `bridge' and `corner' phases, whose features reflect the details of the compact space. As concrete examples we discuss twisted compactifications of 4d ${\cal N}=4$ SYM on $T^2$, $S^2$ and hyperbolic Riemann surfaces.}

\keywords{}

\date{\today}

\begin{document}

\maketitle

\section{Introduction}

The concept of quantum entanglement has provided crucial insight into various areas of physics, including quantum field theory and gravity. In particular, entanglement entropy has played a central role in quantitatively  describing a quintessential property of quantum field theories - renormalization group flows. 

Monotonicity theorems stand as the crowning achievement in the description of renormalization group (RG) flows in quantum field theory. They formalize the notions of counting degrees of freedom and of RG flows as a coarse graining process. The paradigmatic $c$-theorem establishing an RG monotone in 2d was proved by Zamolodchikov \cite{Zamolodchikov:1986gt}, the a-theorem in 4d by Komargodski-Schwimmer \cite{Cardy:1988cwa,Komargodski:2011vj}, and the 3d F-theorem by Casini-Huerta \cite{Jafferis:2010un,Klebanov:2011gs,Jafferis:2011zi,Casini:2012ei}. 

The entanglement entropy (EE) point of view on RG flows has provided particularly useful perspectives and has allowed for  a uniform proof of the above mentioned monotonicity results in $d=2,3,4$ using the strong subadditivity property of entanglement entropy \cite{Casini:2006es,Casini:2017roe,Casini:2017vbe}. Recall that in quantum field theory we are often interested in characterizing the quantum entanglement between a region and its complement. The region that leads to these impressive results is the maximally spherical spatial region in a given dimension. 
    
A particular type of RG flows that has received less attention in the EE context pertains to RG flows across dimensions. Namely, RG flows triggered by compactifications of a $D$-dimensional CFT, which is the UV fixed point, on a $(D-d)$-dimensional compact space, such that the IR fixed point is a $d$-dimensional CFT. The existence of such flows in field theory follows from the standard paradigm of effective field theory; below the energy scale corresponding to the compactification manifold  the theory becomes lower dimensional. The lack of full Lorentz invariance complicates quantum field theory studies. Conveniently,  there is a plethora of  supergravity solutions which precisely describe  RG flows across dimensions  \cite{Maldacena:2000mw,Acharya:2000mu,Gauntlett:2000ng,Gauntlett:2001qs,Gauntlett:2001jj,Benini:2013cda,Benini:2015bwz,Bobev:2017uzs}, thus providing a fertile ground to quantitatively explore them holographically.
    
Motivated by the success of the entropic approach in understanding the behavior of central charges and the ubiquity of holographic solutions describing RG flows across dimensions, \cite{Bea:2015fja,GonzalezLezcano:2022mcd} began explorations  of entanglement across dimensions with the aim of identifying potential counting functions. In this manuscript we  study the behavior of the entanglement entropy along RG flows across dimensions more broadly, with a focus on the interplay between the topology of the compactification manifold and phases of the entanglement entropy. 

In particular, we track entropies which interpolate between the EE of a $(D-1)$-dimensional spherical region (at a fixed time) in the UV and the EE of a $(d-1)$-dimensional spherical region in the IR. Due to the non-trivial (compactified) CFT geometry, the regions interpolating between these two end points along the flow are not spherically symmetric.
This complicates the computation of the associated EE's (which is harder compared to the regions considered in \cite{Bea:2015fja,GonzalezLezcano:2022mcd}). However, interpolating between the maximally spherical entangling regions in the respective dimensions in the UV and IR provides natural connections to the $c$-functions at the fixed points, and turns out to capture the topology of the compactification manifold in an interesting way through corner constributions along the flow.
The minimal surfaces computing the EE's of interest holographically are characterized by partial differential equations. We construct them numerically using the Surface Evolver program \cite{Brakke,Brakke1992} which constructs approximations of increasing quality via iterative triangulations. 
    
The rest of the manuscript is organized as follows. In section \ref{Sec:EntropyGeneral} we recall relevant aspects of entanglement entropy and its holographic computation, and reproduce known results, including corner contributions, using our numerical approach. Section \ref{sec:across_dim} discusses aspects of entanglement in flows across dimensions in a simple holographic model, and we demonstrate the ubiquity of a ``bridge" phase. In section \ref{sec:evolver_results} we discuss the EE evolution in twisted compactifications of ${\cal N}=4$ SYM on $\mathbb{T}^2$, $\mathbb{S}^2$ and higher-genus Riemann surfaces. We conclude in section \ref{Sec:Conclusions}.  Technical details regarding the Surface Evolver and supergravity solutions describing compactifications of $\mathcal N=4$ SYM are given in appendices.

\section{Entanglement Entropy from the Surface Evolver}\label{Sec:EntropyGeneral}

The entanglement entropy associated with a spatial region $\cal A$ in a quantum field theory is the von Neumann entropy of the reduced density matrix, obtained by tracing over the degrees of freedom in the complement of $\cal A$  \cite{Srednicki1993,Casini2009}:
\begin{equation}
    S_{EE}({\cal A}) = -{\rm Tr} \rho_{\cal A} \log \rho_{\cal A}.
\end{equation}
This quantity is divergent. A more proper definition in terms of  operator algebras is described in the exposition  \cite{Witten:2018zxz}. In this manuscript we take this divergent quantity as it is and where appropriate define finite quantities from linear combinations.

Our focus will be on families of EE's interpolating between the EE's for maximally spherical regions in UV and IR CFTs of different dimensions. Within a CFT on flat space, the EE for a spherical region depends only on the dimensionless ratio of the radius of the entangling region $R$ and the short-distance cutoff $\epsilon$.
Specifically for a $D$-dimensional CFT on $\RR^{1,D-1}$, the EE for a $(D-1)$-dimensional spatial ball of radius $R$, $B^{D-1}$, takes the form 
\begin{equation}
\label{Eq:EE-General}
    S_\text{EE}(R;B^{D-1},\epsilon)=\mu_{D-2}\fft{R^{D-2}}{\epsilon^{D-2}}+\mu_{D-4}\fft{R^{D-4}}{\epsilon^{D-4}} + \ldots +
    \begin{cases}
	(-1)^{\frac{D}{2}-1} \, 4A \, \log (R/\epsilon) &(D~\text{even})\\
	(-1)^{\frac{D-1}{2} }\,\,F & (D~\text{odd})
    \end{cases}~.
\end{equation}
The central charge of the CFT can be extracted from the universal cut-off independent part, namely $A$ in even dimensions and $F$ in odd dimensions. 

Our interest will be in more general regions which interpolate between spherical regions of different dimensions. 
The corresponding EE's take a more complicated form.
In particular, we will encounter entangling regions with singular points or loci, such as corners, points with conical geometry or higher dimensional analogs. In those cases, additional divergent terms appear in the expansion of the EE. We will focus on 4d CFTs, for which the entangling region has a 2d boundary. In this case, boundary points with locally conical geometry contribute a $\log^2{\epsilon}$ term, and 1-dimensional extended singularities (``folds'') contribute a $1/\epsilon$ term \cite{Myers2012}.

The quantum field theories we will consider are D-dimensional in the UV and admit holographic descriptions in terms of supergravity solutions which are asymptotically $\rm AdS_{D+1}$. The entanglement entropy of a region $\cal A$ is then computed, following the Ryu-Takayanagi prescription \cite{Ryu:2006bv,Ryu:2006ef,Nishioka:2009un}, by finding a minimal surface $\gamma_{\cal A}$ in the bulk which is anchored at the conformal boundary of $\rm AdS_{D+1}$ on the boundary of $\cal A$, $\partial \gamma_{\cal A}=\partial\cal A$. The EE then is 
\begin{equation}
    S_{EE}({\cal A}) = \frac{\text{Area}(\gamma_{\cal A})}{4G_N^{D+1}},
\end{equation}
where $G_N^{(D+1)}$ is Newton's constant in $D+1$ dimensions. The area of the surface will be divergent and can be regularized, for example, by introducing a cut-off close to the conformal boundary.
This reproduces the structure in eq.~(\ref{Eq:EE-General}) for spherical entangling regions.

The problem of finding minimal surfaces with given boundary conditions, known as Plateau's Problem, is a well established problem in mathematics dating back to Lagrange in 1790, tackled experimentally with soap bubbles by Plateau in the 1830's and with its first general solutions found in 1930 by Douglas \cite{10.2307/1989472} and Rad\'o \cite{10.2307/1968237}. In this work we use the \textit{Surface Evolver} \cite{Brakke,Brakke1992}, a program that constructs polygonal surfaces using triangulations to approximate minimal surfaces for given boundary conditions. The Surface Evolver has been previously used to study entanglement entropy in holographic duals for 3D CFTs \cite{Fonda2015,Seminara2017}. We will use it for entangling regions with sufficient symmetry in the holographic dual geometries corresponding to RG flows across dimensions. To set the stage, we now illustrate and validate the results obtained by this approach for the EE of two simple example regions.

\subsection{Infinite cylinder in AdS$_5$} \label{sec:inf_cylinder}

The first example is an infinite cylinder in a 4d CFT on $\RR^{1,3}$. As holographic dual we take (unit radius) $\rm AdS_5$ with cylindrical coordinates for the spatial part of the CFT geometry, 
\be \label{eq:AdS5-metric}
ds^2_{AdS_5}=\frac{1}{z^2}\left(-dt^2+dz^2+dx^2+dr^2+r^2d\phi^2\right),
\ee
The cylindrical region $\cal A$ corresponds to $r\leq R$. 
The bulk minimal surface anchored on the boundary of the cylinder, $r=R$, can be defined by a function $z(r)$ alone, or equivalently $r(z)$ alone.
For the numerical implementation, a small $\epsilon$ of the order of 0.01 times the length scale of the entangling region is chosen and the boundary conditions  are imposed at $z=\epsilon$ rather than $z=0$. This modified prescription does not change the universal term of the EE, but may affect non-universal terms. 
The induced metric on the surface (at a constant time) is
\be 
ds_{ind}^2=\frac{1}{z^2}\left(dx^2+r(z)^2d\phi^2 + (1+r'(z)^2)dz^2\right).
\ee
The area is given by
\begin{equation}
    \text{Area}(\gamma_{\cal A}) =  2 \int_0^{2\pi}d\phi\int_{-h/2}^{h/2}dx\int_\epsilon^{z_0}dz \sqrt{\det g_{ind}} =  4 \pi h\int_\epsilon^{z_0}\frac{r(z)}{z^3}\sqrt{1+r'(z)^2}dz\,,
    \label{cylinder_area}
\end{equation}
where $z_0$ is the solution of $r(z)=0$, and $h$ is a cutoff for the length of the cylinder.

The minimal surface can be found numerically using the  curve length minimization procedure described in app.~\ref{App:SurfaceEvolver}, in  which the function $r(z)$ is approximated by a series of line segments and then iteratively refined. Eq. (\ref{cylinder_area}) is implemented as the functional
\begin{equation}
    \mathcal{F}[C]=2 \pi h\oint_C \frac{r}{z^3}ds,
    \label{cylinder_func_actual}
\end{equation}
in the Surface Evolver, where the integral is taken over all segments of the curve $C$ with respect to Euclidean distance in $(r,z)$ space.

\begin{figure}
     \centering
     \begin{subfigure}[h!]{0.29\textwidth}
         \centering
         \includegraphics[width=0.98\textwidth]{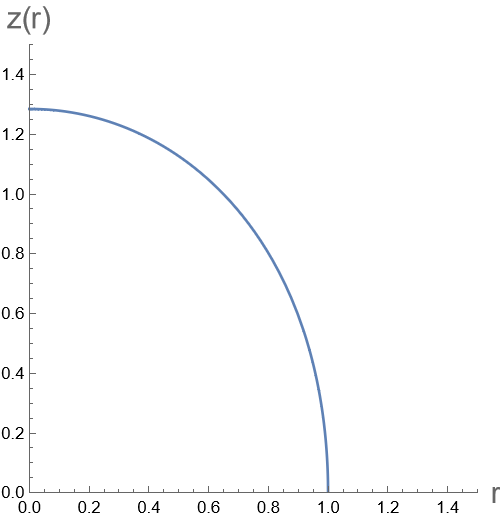}
     \end{subfigure}
     \quad
     \begin{subfigure}[h!]{0.55\textwidth}
         \centering
         \vspace{10 pt}
         \includegraphics[width=\textwidth]{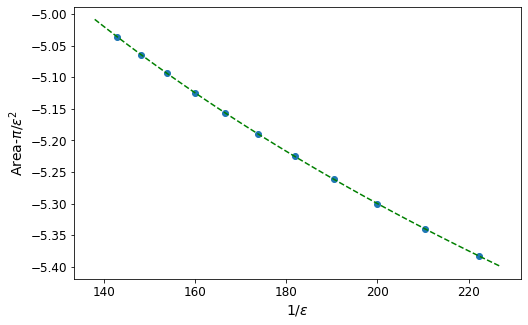}
     \end{subfigure}
        \caption{Left: Minimal surface for an infinite cylinder with $R=1$, obtained numerically as solution to equation (\ref{cylinder_EL}). Right: Areas for 11 different $\epsilon$ as determined by the Surface Evolver (after subtracting the $1/\epsilon^2$ contribution), and the best fit curve based on (\ref{cylinder_exp}).}
        \label{fig:cylinder_data}
\end{figure}

With $R$ and $h$ fixed, the minimal area is determined for various values of $\epsilon$. This numerical data is then used to fix the coefficients of the $\epsilon$ expansion
\begin{equation}
    \text{Area}(\gamma_{\cal A}) = c_2\frac{hR}{\epsilon^2} + c_l\frac{h}{R}\log\left(\frac{R}{\epsilon}\right) + c_0\frac{h}{R} + \mathcal{O}(\epsilon^2),
    \label{cylinder_exp}
\end{equation}
which is compatible with $h$ only appearing as an overall factor and the dependence on $R$ following from dimensional analysis.  We have that  $c_2=\pi$ since the coefficient of the leading term is given by the area of the entangling region devided by $D-2$ for a CFT$_D$. 
A least squares fit based on the Surface Evolver data for $R=h=1$, shown in Fig. \ref{fig:cylinder_data}, leads to
\begin{equation}
    c_l=-0.7854045 \hspace{1 cm}\text{and}\hspace{1 cm} c_0=-1.1385438.
    \label{cyl_results}
\end{equation}
To improve the quality of the fit we included terms up to order $\epsilon^2$. With $c_2$ fixed, this leaves three parameters to be determined.

To assess the accuracy of the fit, we also determine $c_l$ analytically. The Euler-Lagrange equation resulting from (\ref{cylinder_area}) is
\begin{equation}\label{cylinder_EL}
    (z+3rr')(1+(r')^2)-zrr''=0,
\end{equation}
which constrains the expansion for $r(z)$ for small $z$ to
\be\label{Eq:r-z-expansion}
r(z)=R-\frac{1}{4R}z^2+\frac{1}{32R^3}z^4\log z+\mathcal{O}(z^4).
\ee
Using this expansion in (\ref{cylinder_area}) yields 
\begin{equation}\label{cyl_actual_exp}
    \text{Area}(\gamma_{\cal A}) = \frac{\pi R L}{\epsilon^2} - \frac{\pi L}{4 R}\log\left(\frac{R}{\epsilon}\right) + \mathcal{O}(\epsilon^0).
\end{equation}
The numerical value for $c_l$ in (\ref{cyl_results}) agrees with the analytical result (\ref{cyl_actual_exp}) to within $10^{-5}$. The general form of this coefficient was derived in \cite{Solodukhin:2008dh} using field-theoretic methods, which shows that the $c$ central charge appears in the $\log$ coefficient for the cylindrical entangling region, not the pure $a$ central charge which appears for a spherical entangling region in (\ref{Eq:EE-General}).

\subsection{Spindle with conical singularities} \label{subsec:singular}

\begin{figure}
     \centering
         \includegraphics[width=0.2\textwidth]{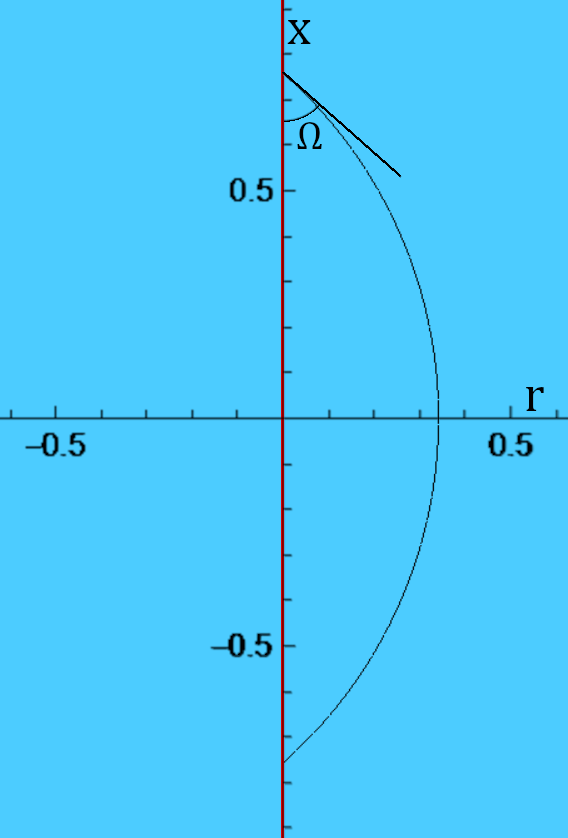}
     \qquad
         \includegraphics[width=0.5\textwidth]{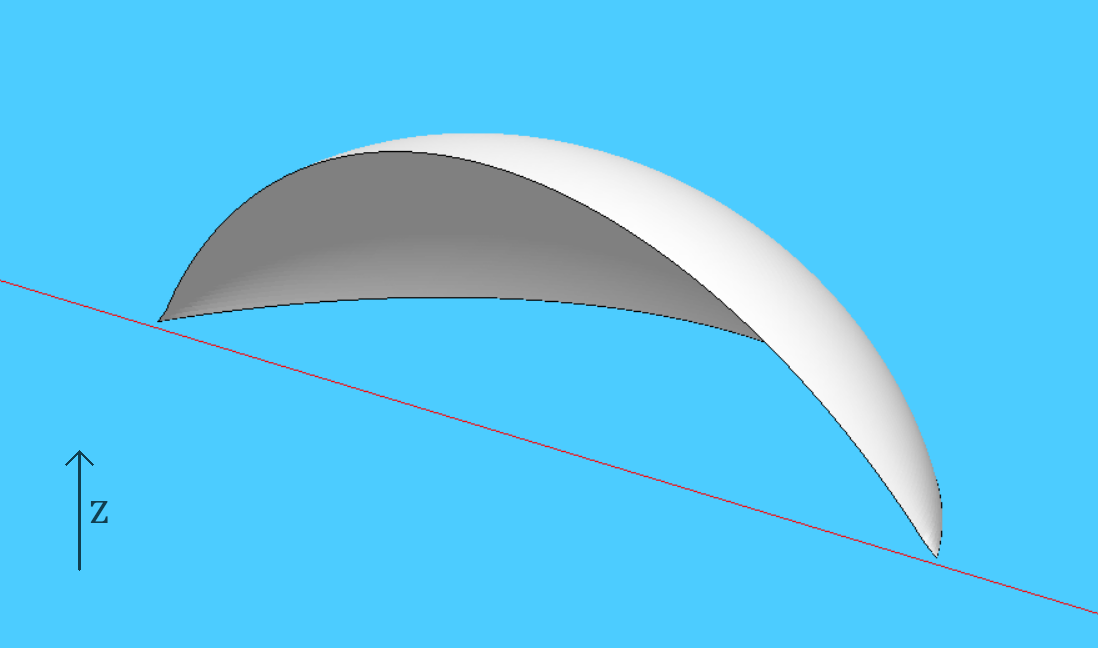}
        \caption{Left: Profile of the region described in section \ref{subsec:singular} with $\Omega\approx 0.85$. The full 3D entangling region is formed by rotating the curve around the red vertical line. Right: The minimal surface $z(r,x)$ in the bulk, as computed by the Surface Evolver, with the holographic coordinate $z$ pointing upwards.}
        \label{fig:corners}
\end{figure}

As a simple example of a region with conical singularities we consider the spindle-shaped region defined by rotating a circular arc around the line defined by its endpoints (Fig. \ref{fig:corners}). This produces an axially symmetric region whose boundary contains two singular points.

If we introduce cylindrical coordinates $(r,\theta,x)$ for the CFT in the $\rm AdS_5$ metric in (\ref{eq:AdS5-metric}), with $x$ along the axis of revolution, the minimal surface is defined by $z=z(r,x)$. Thus, the problem reduces to finding a two-dimensional surface in $(r,x,z)$ space. The area is
$$2\pi \iint_\Sigma\frac{r}{z^3}\sqrt{1+z_r^2+z_x^2}drdx,$$
and we numerically approximate it by adding contributions from all facets of a finite triangulation. The coefficients of the $\epsilon$ expansion can be extracted as in the case of the cylinder before, although an additional $\log^2(\epsilon)$ term must be included due to the singular points. This term has an analytical form obtained in \cite{Myers2012},
\begin{equation}
    -\frac{\pi}{8}\cos(\Omega)\cot(\Omega)\log^2(\epsilon),
    \label{corner_log2}
\end{equation}
where $\Omega$ is the angle from the axis to the surface of the conical singularity. As shown in Table \ref{tab:corner_log2}, the results obtained using the Surface Evolver are in close agreement with the analytical values. In the following sections, this type of divergent term will appear generically as contribution to the entanglement entropy as we connect spherical regions across dimensions.

\begin{table}[h!]
\begin{center}
\begin{tabular}{ |c||c|c|c|c|  }
 \hline
 $\Omega$ & Analytical & Numerical & Difference & \% Error\\
 \hline
 $\pi/10$ & $1.1495$ & $1.1432$ & $6.3\times 10^{-3}$ & $0.55 \%$\\
 $\pi/6$ & $0.5890$ & $0.5795$ & $9.6\times 10^{-3}$ & $1.7 \%$\\
 $\pi/4$ & $0.2777$ & $0.2760$ & $1.7\times 10^{-3}$ & $0.62 \%$\\
 $2\pi/5$ & $0.03943$ & $0.03934$ & $8.9\times 10^{-5}$ & $0.23 \%$\\
 \hline
\end{tabular}
\end{center}
\caption{$\log^2\epsilon$ coefficient for different opening angles $\Omega$. The numerical values were computed by curve fitting the area of the surface described in section \ref{subsec:singular}.}
\label{tab:corner_log2}
\end{table}

\section{EE across dimensions: caps, bridges and corners}
\label{sec:across_dim}

In CFTs the universal term in the EE of a spherical region encodes the central charge or free energy (eq.~(\ref{Eq:EE-General})). For RG flows within the same dimension, spherical regions of varying size $R$ can be used to define $c$-functions, which connect the central charges at the UV and IR fixed points montonically and track the effective number of degrees of freedom along the flow. Suitable derivatives of the EE with respect to $R$ extract the universal coefficients at the fixed points and interpolate between them along the flow where multiple scales compete (see e.g.\ \cite{Liu:2012eea,Casini:2017vbe}).
In this section we describe new aspects in the evolution of the EE for spherical regions of increasing size for RG flows across dimensions.

We are interested in $D$-dimensional CFTs partially compactified on $\RR^{1,d-1}\times M_{D-d}$, such that a $d$-dimensional CFT emerges in the IR. There is a plethora of supergravity solutions with known field theory duals which describe such RG flows across dimensions for various choices of $M_{D-d}$, including spheres, tori and hyperbolic manifolds \cite{Maldacena:2000mw,Acharya:2000mu,Gauntlett:2000ng,Gauntlett:2001qs,Gauntlett:2001jj,Benini:2013cda,Benini:2015bwz,Bobev:2017uzs}. 
The solutions interpolate between $\rm AdS_{D+1}$ in the UV and $\rm AdS_{d+1}\times M_{D-d}$ in the IR. The metric takes the general form 
\begin{equation}
	ds^2_{D+1}=e^{2f(z)}\left(\eta_{\mu\nu}dx^\mu dx^\nu+ dz^2\right) + e^{2g(z)} g_{ij}(y)dy^i dy^j,\label{flow:across}
\end{equation}
where $z$ is the holographic coordinate corresponding to the QFT energy scale and the functions $f$ and $g$ describe the interpolation. 
 The coordinates $y^i$ describe the compact manifold $M_{D-d}$,  $\mu,\nu=0,.., d-1$ describe the geometry of the IR CFT$_d$, and the geometry of the CFT$_D$ in the UV is described by $(x^\mu, y^i)$. 
The compactification breaks $D$-dimensional Lorentz symmetry. Concrete examples will be discussed in sec.~\ref{sec:evolver_results}; the discussion here will be qualitative.

To interpolate between the UV and IR central charges using entanglement entropies for such flows across dimensions, we start with the topology of the natural entangling regions at the UV and IR fixed points. In the UV, the entangling region should be a $(D-1)$-dimensional ball $B^{D-1}$, while in the IR it should wrap the compact space entirely and remain spherical only in the non-compact directions, i.e.,  $B^{d-1}\times M_{D-d}$. In order to define an interpolating function connecting the central charges using EE, we should thus consider a continuous evolution of entangling regions that transitions between the two topologies. Crucially, this necessitates topology change along the flow. We will now discuss this for a simple toy model.

\begin{figure}
    \centering
    	\begin{tikzpicture}
		\foreach \i in {0,...,3} \draw[dashed] (\i,0.4) -- (\i,3.6);
		
		\draw (0.5,2) circle (5pt);
		\draw (1.5,2) circle (5pt);
		\draw (2.5,2) circle (5pt);
	\end{tikzpicture}
\hskip 20mm
	\begin{tikzpicture}
		\draw (0.5,2) circle (15pt);
		\draw (1.5,2) circle (15pt);
		\draw (2.5,2) circle (15pt);
		
		\draw[white,fill=white]  (-0.25,1.82) rectangle (3.2,2.18);
		
		\foreach \i in {0,...,3} \draw[dashed] (\i,0.4) -- (\i,3.6);
	\end{tikzpicture}
\hskip 20mm
	\begin{tikzpicture}
		\draw (0.5,2) circle (30pt);
		\draw (1.5,2) circle (30pt);
		\draw (2.5,2) circle (30pt);
	
		\draw[white,fill=white]  (-0.7,1.08) rectangle (3.7,2.92);
		
		\foreach \i in {0,...,3} \draw[dashed] (\i,0.4) -- (\i,3.6);
	
	\end{tikzpicture}

    \caption{Entangling regions of varying radius on $\mathbb{R}\times S^1$. The $S^1$ direction runs horizontally and its identification are denoted by the dashed vertical lines. Figure reproduced from \cite{GonzalezLezcano:2022mcd}}.
    \label{fig:boundary_evolution}
\end{figure}
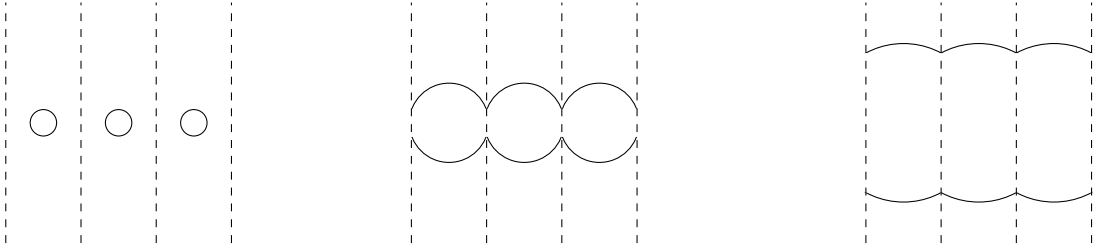

Fig.~\ref{fig:boundary_evolution} shows an example of entangling regions of increasing size for a QFT$_3$ with spatial geometry $\mathbb{R}\times S^1$, where the $S^1=\RR/\ZZ$ direction runs horizontally, with fundamental domains separated by dashed vertical lines.  The entangling regions shown are defined by
\begin{align}\label{eq:region-def}
\{(x,y)\in\mathbb{R}\times S^1|\;d(x,y)\leq R\}\,,
\end{align}
where $d$ is the distance to a reference point in $\RR\times S^1$.
For small $R$ the entangling region is spherical in the two-dimensional sense, corresponding to the geometry of the higher-dimensional UV QFT. But when $2R$ exceeds the size of the $S^1$ the entangling region starts to partially wrap the compact direction. At that point the spherical regions in adjacent fundamental domains start to overlap (Fig.~\ref{fig:boundary_evolution} center),  and the entangling region develops corners as a result. As $R$ increases further, the corners smooth out and the region approaches the product form $B^1\times S^1$. In the limit of large $R$ it wraps the $S^1$ entirely and becomes spherical in the one-dimensional sense, corresponding to the geometry of the lower-dimensional IR QFT.

We note that the area a of the entangling surface (i.e.\ the boundary of the entangling region) also shows non-trivial behavior as $R$ is increased. For small $R$ (Fig.~\ref{fig:boundary_evolution} left), the entangling surface is a circle with circumference $2\pi R$, as appropriate for a spherical region in the higher-dimensional UV geometry. This linear growth continues up to the point where $R$ is half the size of the $S^1$. At this point the entangling surface has maximal area. Upon further increasing $R$, the area of the entangling surface decreases (e.g.\ in the transition from the center to the right panel in Fig.~\ref{fig:boundary_evolution}), and it asymptotically approaches twice the length of the $S^1$. The area of the entangling surface sets the leading divergence in the RT surfaces to be discussed below, and the general behavior is  $R^{D-2}$ for small $R$ and $R^{d-2}$ for large $R$.

Entanglement entropy in flows across dimensions was already studied in \cite{GonzalezLezcano:2022mcd}, but only for regions that fully wrap the compact part of the geometry. For the example above these wrapping regions take the form of a product of an interval and $S^1$,
$$\{(x,y)\in\mathbb{R}\times S^1|\;|x|\leq R\}.$$ 
They have the same large $R$ limit as the example in Fig.~\ref{fig:boundary_evolution} and capture the IR central charge. But they differ for finite $R$ and never become fully spherical for small $R$. As a result, the universal term of the EE does not approach the monotonic central charge of the higher-dimensional theory in the UV, but rather a linear combination of, e.g.,\ the $a$ and $c$ type central charges. This makes the process in fig.~\ref{fig:boundary_evolution} better suited to capture the behavior of the UV theory while also capturing the IR central charge and providing a natural connection.

\begin{figure}
     \centering
     \begin{subfigure}[h!]{0.4\textwidth}
         \centering
         \includegraphics[width=\textwidth]{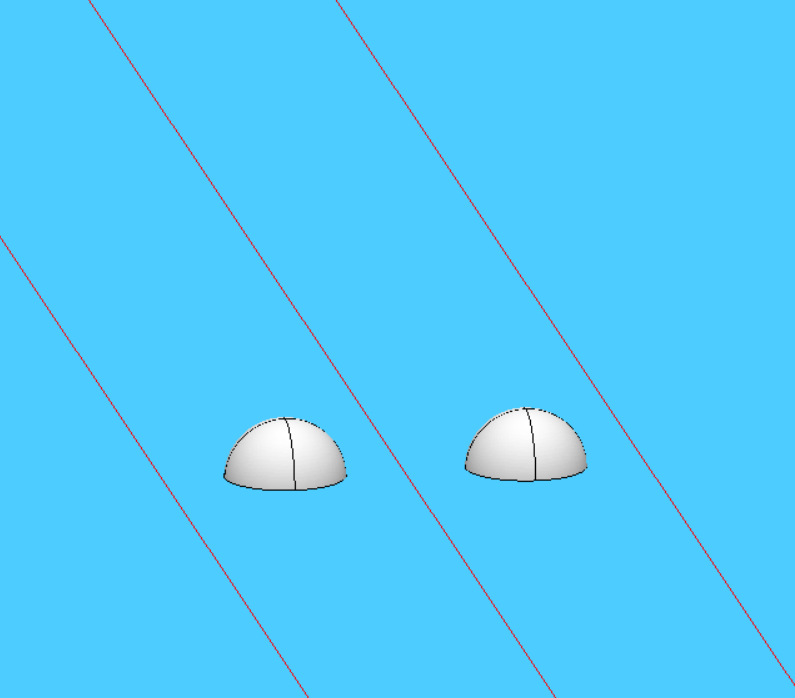}
         \caption{R=0.25}
     \end{subfigure}
     \qquad
     \begin{subfigure}[h!]{0.4\textwidth}
         \centering
         \includegraphics[width=\textwidth]{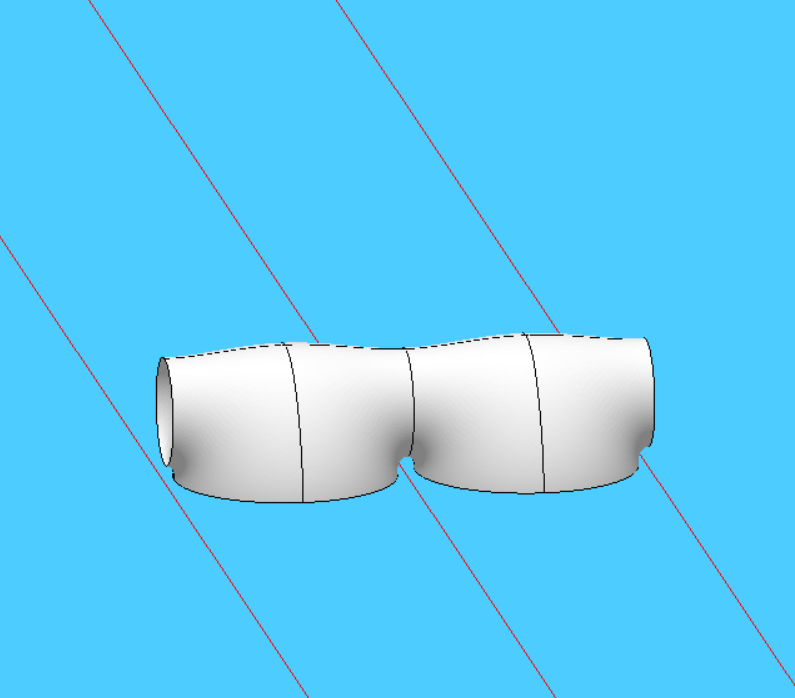}
         \caption{R=0.46}
     \end{subfigure}
     \vskip\baselineskip
     \begin{subfigure}[h!]{0.4\textwidth}
         \centering
         \includegraphics[width=\textwidth]{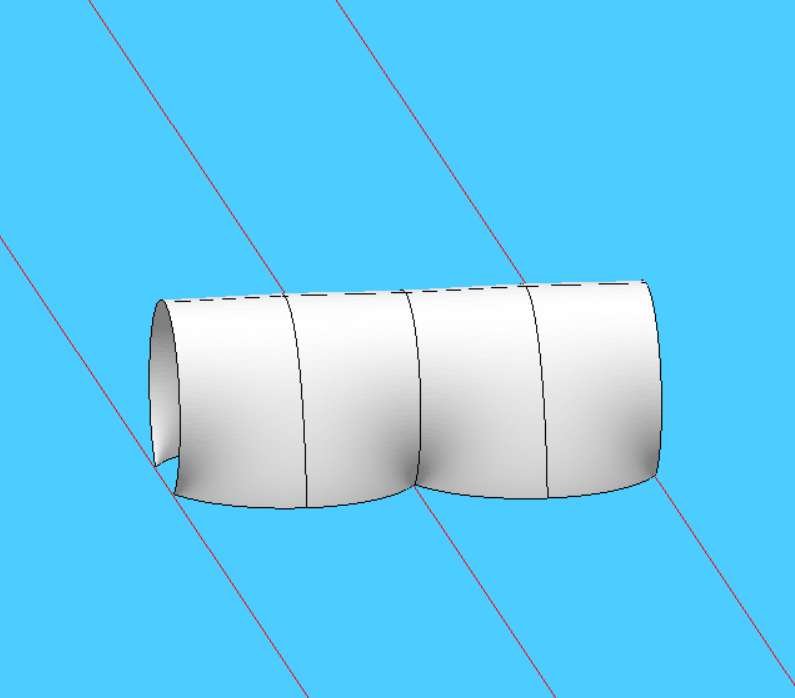}
         \caption{R=0.55}
     \end{subfigure}
     \qquad
     \begin{subfigure}[h!]{0.4\textwidth}
         \centering
         \includegraphics[width=\textwidth]{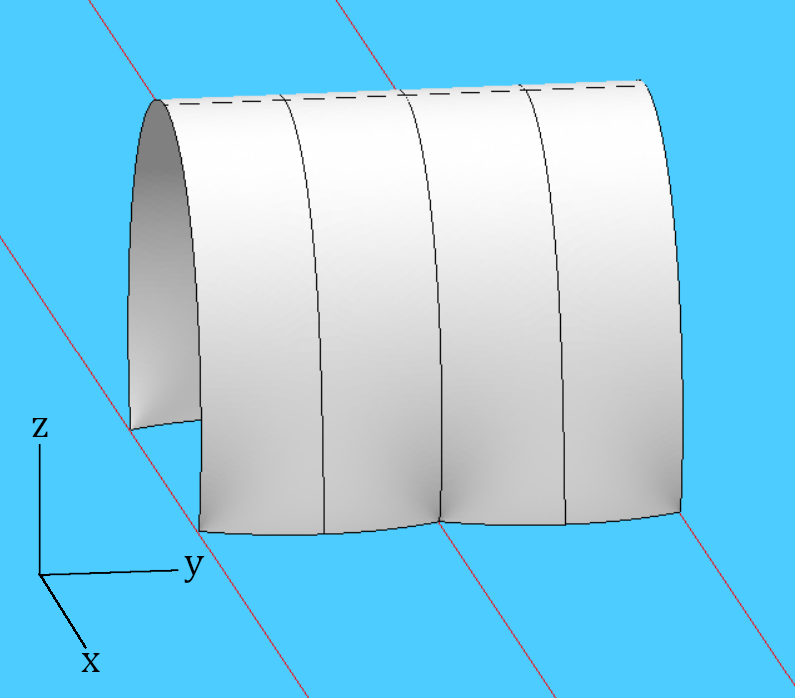}
         \caption{R=1.00}
     \end{subfigure}
        \caption{Example minimal surfaces for the process in Fig. \ref{fig:boundary_evolution}. The CFT has spatial geometry $\mathbb{R}\times S^1$ and the bulk is taken as AdS$_4$ with periodic identification. The red lines, at unit separation in the $y$ direction, are identified. For small regions (a), the minimal surface has the shape of a spherical cap. As the region approaches the size of the compact space, a topology transition occurs and ``bridges''form between neighboring domains (panel (b)). In (c), the diameter of the region surpasses one and the EE picks up corner contributions. As the region continues to grow, these contributions decrease (d).}
        \label{fig:topology_evolution}
\end{figure}

Fig.\ \ref{fig:topology_evolution} shows examples of bulk 
RT surfaces anchored on the entangling surfaces in Fig.\ \ref{fig:boundary_evolution}. These were produced by considering the Poincare metric for AdS$_4$ with periodicity introduced in one of the spatial coordinates:
\begin{equation} \label{ads4_metric}
    ds^2 = \frac{L^2}{z^2}\left(-dt^2+dz^2+dx^2+dy^2\right),\hspace{40 pt} y\sim y+\ell\,.
\end{equation}
This metric does not become asymptotically AdS$_3$ for $z\to \infty$, but it will be shown later that it is a good model for certain aspects in RG flows across dimensions. In the figure, we take $L=\ell=1$. For small regions the minimal surface is a hemisphere, unchanged from the case of pure AdS$_4$. However, at some point as the size of the region approaches the size of the compact direction, the minimal surface changes topology, (b). At this phase transition, the surface develops a bridge-like component that causes it to fully wrap the periodic direction for some range of $z$. Going forward, (a) and (b) will be referred to as the ``cap'' and ``bridge'' configurations, respectively. As $R$ increases further, the gap under the bridge collapses and the surface develops sharp ``corners'' at the boundary, (c). This corresponds to the transition in Fig.~\ref{fig:boundary_evolution}. For large regions the corners smooth out, Fig.\ \ref{fig:topology_evolution}(d). One of the main results of this manuscript is to provide evidence, in a holographic context, for the universality of this sequence of entanglement phases: caps, bridges and corners. 

The topology change and the appearance of corners both create phase transitions in the EE along the RG flow. The corners lead to new divergences in the EE, which take a universal form and are well understood \cite{Myers2012}. We will present a generic argument for the existence of a bridge phase before the onset of corners in sec.~\ref{sec:bridge_argument}, and a quantitative discussion of the transition to the bridge phase and the onset of corners will be given in sec.~\ref{sec:evolver_results}.

\subsection{Bridges before corners} \label{sec:bridge_argument}

In Fig. \ref{fig:topology_evolution}, a transition occurs between the cap and bridge configurations for a particular $R$ close to but less than half the length $\ell$ of the compact direction. We now present an argument for why this is the general behavior.

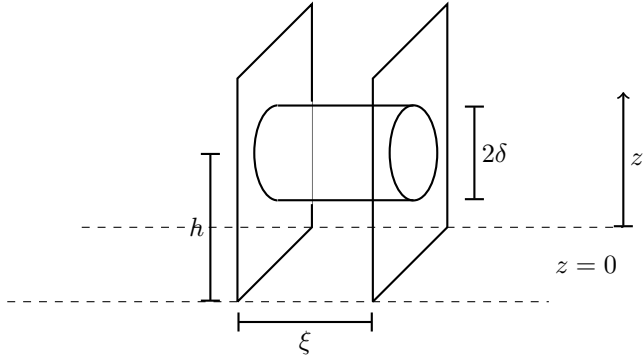
\begin{figure}
    \centering
    \begin{tikzpicture}[scale=0.9]
	\draw[thick] (-0.6,-2.2) -- (-0.6,1.1) -- +(1.1,1.1) -- (-0.6+1.1,-2.2+1.1) -- (-0.6,-2.2);
	\draw[thick] (0,0) ellipse (10pt and 20pt);
	\draw[fill=white,white] (0,-0.75) rectangle (0.5,0.75);
	\draw[thick] (2,0) ellipse (10pt and 20pt);
	\draw[thick] (0,0.7) -- +(2,0);
	\draw[thick] (0,-0.7) -- +(2,0);
	\draw[thick] (2-0.6,-2.2) -- (2-0.6,1.1) -- +(1.1,1.1) -- (2-0.6+1.1,-2.2+1.1) -- (2-0.6,-2.2);
	
	\draw [|-|,thick] (-0.6,-2.5) -- +(2,0);
	\node at (-0.6+1,-2.8) {\small $\xi$};
	
	\draw [|-|,thick] (2.9,-0.7) -- (2.9,0.7);
	\node at (3.2,0) {\small $2\delta$};

 	\draw[|-|,thick] (-1,-2.2) -- (-1,0);
	\node at (-1.2,-1.1) {\small $h$};

	\draw[dashed] (-4,-2.2) -- +(8,0);
	\draw[dashed] (-4+1.1,-2.2+1.1) -- +(8,0);
	\node at (4+0.55,-2.2+0.55) {\small $z=0$};
	\draw [|->,thick] (4+1.1,-2.2+1.1) -- +(0,2);
	\node at (4+1.1+0.2,-2.2+1.1+1) {\small $z$};
    \end{tikzpicture}
    \caption{When the caps in fig.~\ref{fig:topology_evolution}(a) are grown to $R\lesssim \ell/2$, the surfaces in neighboring fundamental domains almost touch. Close to the point where they would touch for $R=\ell/2$, the two surfaces can be approximated as two parallel planes, as shown here. The formation of a small cylindrical bridge as shown then decreases the area, leading generically to the phase in fig.~\ref{fig:topology_evolution}(b).\label{fig:bridge_argument}}
\end{figure}

As the caps in fig.~\ref{fig:topology_evolution}(a) grow, the surfaces in neighboring fundamental domains eventually almost touch. Close to the point where they would eventually touch, for $R$ sufficiently close to $\ell/2$ and for small $z$, the two cap surfaces can be approximated as two parallel planes with a small separation $\xi$. 
Now consider adding a small cylindrical bridge of radius $\delta$, with $\delta$ much smaller than the height $h$ where the bridge is located. 
This is illustrated in Fig \ref{fig:bridge_argument}. 
Under the above approximation, the difference in area is
$$A_\text{bridge} - A_\text{no bridge}\approx\frac{1}{h^{d-2}}(2\pi\delta\xi - 2\pi\delta^2)=\frac{1}{h^{d-2}}2\pi\delta(\xi-\delta).$$
Therefore, adding a bridge with $\delta>\xi$ lowers the area. For $R$ sufficiently close to $\ell/2$, the capped configuration cannot be minimal.

Let us justify the above approximations. Since the curvature scale of the surface is approximately $R$, the size of the gap between the two cap surfaces grows with the AdS radial coordinate $z$ as $z^2/R$. We therefore assume
\begin{align}
    \frac{h^2}{R} &\ll \xi\,, \qquad
    \xi \ll h\,.
\end{align}
The first condition allows us to assume that the surfaces are approximately parallel up to the height of the bridge.
The second inequality is introduced so that $h$ can be assumed constant across the bridge (recalling that $\delta$ can be made comparable to $\xi$ as long as it is still greater). Choosing some large $N$, we can realize these conditions as
$N h^2/R = \xi$ and $N\xi = h$,
which implies $\xi = R/N^3$ and $h=R/N^2$. We note that the above argument does not determine the true minimal surface; it simply shows that a small bridge is favored over no bridge. The approximations do not restrict the actual size of the bridge, which will likely be of order $R$, as in fig. \ref{fig:topology_evolution}(b).
We also note that the argument above is local, in the sense that it does not depend on the geometry of the cap surfaces away from the point where they almost intersect. We therefore expect it to apply for more general compactifications.

\section{Flows from 4d to 2d: ${\cal N}=4$ SYM on $T^2$, $S^2$, $\Sigma_{g>1}$}\label{sec:evolver_results}

In this section we discuss the evolution of the EE for  4d $\mathcal N=4$ SYM compactified on constant curvature Riemann surfaces as a concrete top-down AdS/CFT example, noting that the discussion extends to other similar examples.
The RG flow solutions intepolating between $\rm AdS_5$ in the UV and $\rm AdS_3$ in the IR have been constructed within a consistent truncation of Type IIB supergravity known as the STU model. Details are reviewed in Appendix \ref{App:SugraSolutions}.
We start with torus compactifications, which realize the qualitative discussion of Sec.~\ref{sec:across_dim}, and then discuss $S^2$. For higher-genus surfaces we content ourselves with a qualitative discussion.
 
\subsection{4d $\mathcal N=4$ SYM on $T^2$}\label{Subsect:ThinTorus}

The geometric considerations of Sec.~\ref{sec:across_dim} can be realized in a top-down context by considering 4d $\mathcal N=4$ SYM on a thin torus. We will first discuss how the geometric setup comes about and then analyze the entanglement entropy.

\begin{figure}
    \centering
    \includegraphics[width=0.55\textwidth]{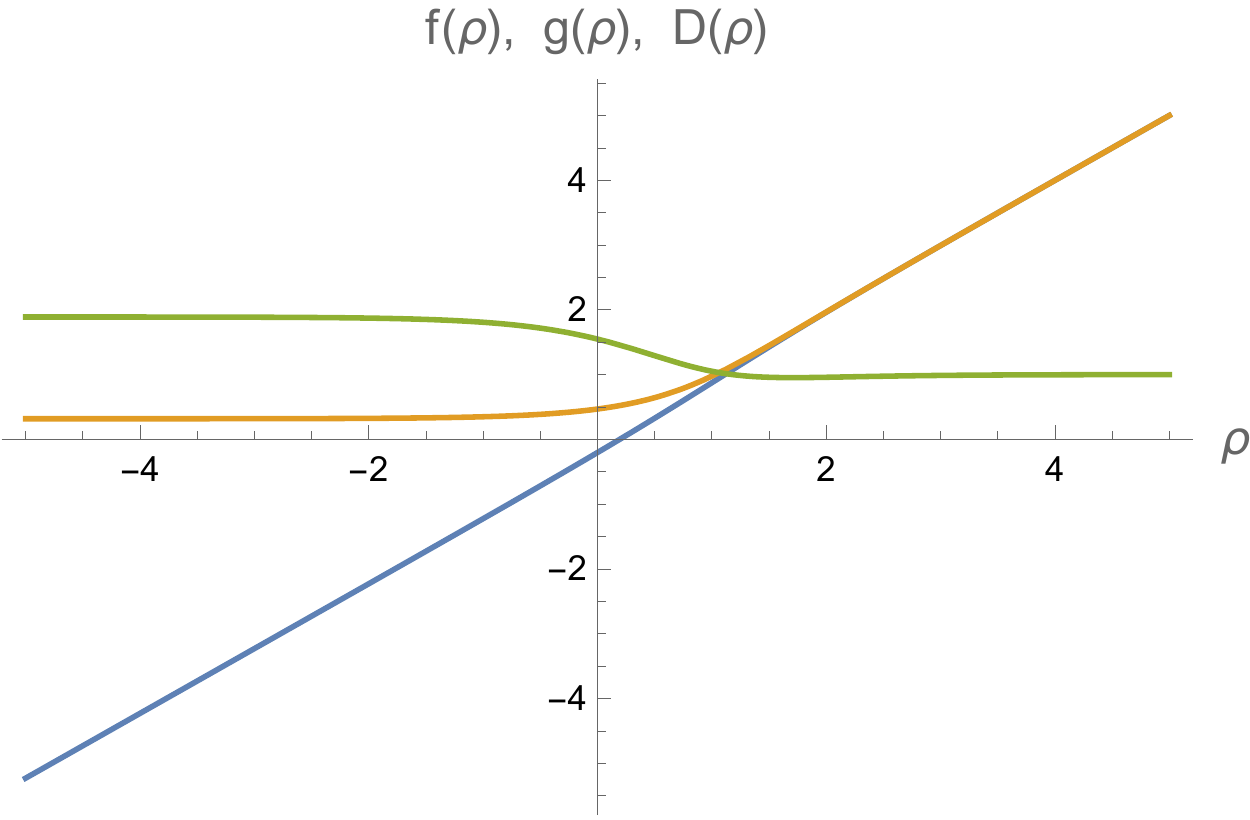}
    \caption{Flow solution for a $\mathbb{T}^2$ compactification, with $f$, $g$ and $D$ in blue, orange, and green, respectively.}
    \label{fig:torus_fgd-main}
\end{figure}

The metric of the holographic RG flow solutions reviewed in Appendix \ref{App:SugraSolutions} takes the form
\begin{equation} \label{5D sugra metric main}
    ds_5^2=e^{2f(\rho)}ds^2_{\RR^{1,1}} + \frac{d\rho^2}{D(\rho)^2} + e^{2g(\rho)}ds^2_{T^2}\; .
\end{equation}
The functions in this Ansatz are determined by the BPS equations in Appendix \ref{App:SugraSolutions}, and an example solution is shown in Fig.~\ref{fig:torus_fgd-main}.
In the UV ($\rho\rightarrow\infty$), $f$ and $g$ are both linear with the same slope, while $D$ is constant. In this regime the geometry approaches $\rm AdS_5$. In the IR ($\rho\rightarrow-\infty$), $f$ is again linear, with $D$ constant, but $g$ now is constant as well. The geometry becomes $\rm AdS_3\times T^2$. The metric functions and the scale of the transition region between these two regimes are determined by the fluxes specifying the twist and the volume of $T^2$.

Since the solution in (\ref{5D sugra metric main}) depends on the geometry of $T^2$ only through the volume, i.e.\ the product of the lengths of the two cycles $\ell_1$ and $\ell_2$, we are free to choose their relative length and consider the case $\ell_1\ll \ell_2$. The energy scale associated with the shorter cycle can then be shifted firmly into the UV region of the geometry, where $\rho$ is large and the solution is very well approximated by $\rm AdS_5$ with one direction compactified. The evolution of the EE for regions whose size, $R$, is comparable to $\ell_1$ is then analogous to Figures \ref{fig:boundary_evolution} and \ref{fig:topology_evolution}.

For the discussion of the EE bridge and corner transition associated with the shorter cycle we can therefore approximate the metric as 
\begin{equation}\label{periodic_ads_metric}
    ds^2 = \frac{L^2}{z^2}\left(-dt^2+dx_1^2+dx_2^2+dx_3^2+dz^2\right), \qquad x_1\sim x_1+\ell_1\,, 
\end{equation}
where $x_1,x_2, x_3$ are the spatial dimensions and $z$ is the holographic direction. From the point of view of an entangling region on the scale of $\ell_1$, the geometry is AdS$_5$ wrapped on a circle. This approximation breaks down once the radius of the entangling region becomes comparable to the scale of the transition region in Fig.~\ref{fig:torus_fgd-main}.  It in particular does not capture the IR limit.

As the radius of the spherical entangling region, $R$, changes, the entangling region will behave analogously to the one in Figure \ref{fig:boundary_evolution}, except that there is an extra spatial direction since we are in $\rm AdS_5$ instead of $\rm AdS_4$. One should therefore think of Figure \ref{fig:boundary_evolution} as rotating around a horizontal axis through the center of the entangling region.
For $2R<\ell_1$ the region will be spherical in the higher-dimensional (UV) sense, while for larger $R$ contiguous copies of the spherical region in adjacent fundamental domains will make contact. Instead of points, the intersections of the copies will be circles. As $R$ is increased further, the region inside the fundamental domain will approximate a wide cylinder.

For the following quantitative analysis of the EE evolution we set $L=\ell_1=1$ for simplicity. For $R<1/2$ and before the bridge transition, the region will be a spherical ``cap" and the EE can be computed analytically.
With spherical coordinates on the spatial part of the AdS metric,
$$ds^2 = \frac{1}{z^2}\left(-dt^2+dz^2+dr^2 +r^2d\Omega^2\right),$$
and the sphere placed at the $r=0$ origin, the area of the minimal surface is
\begin{equation}
    \text{Area}(\gamma_{\cal A})=4\pi\int_\epsilon^{z_0}\frac{r(z)^2}{z^3}\sqrt{1+r'(z)^2}dz.
    \label{sphere_area}
\end{equation}
With the boundary condition $r(0)=R$, the minimal surface would be $r(z)=\sqrt{R^2-z^2}$. However, for the numerics we use $r(\epsilon)=R$, in which case the minimal surface becomes $r(z)=\sqrt{R^2+\epsilon^2-z^2}$. Both choices yield the same leading divergence and universal term, but differ in the constant term. With the boundary condition at $z=\epsilon$, the area is
\begin{equation} \label{sphere exp}
     \text{Area}(\gamma_{\cal A}) = \frac{2\pi R^2}{\epsilon^2} - 2\pi\log\left(\frac{2R}{\epsilon}\right) +\pi + \mathcal{O}(\epsilon^2)\,.
\end{equation}

\begin{figure}
     \centering
     \begin{subfigure}[h!]{0.49\textwidth}
         \centering
         \includegraphics[width=\textwidth]{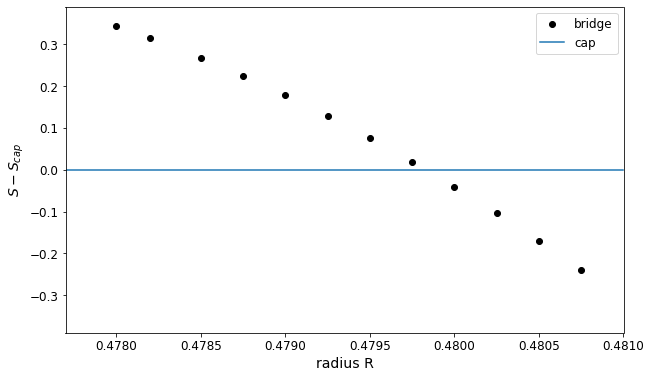}
     \end{subfigure}
     \hspace{0 pt}
     \begin{subfigure}[h!]{0.49\textwidth}
         \centering
         \includegraphics[width=\textwidth]{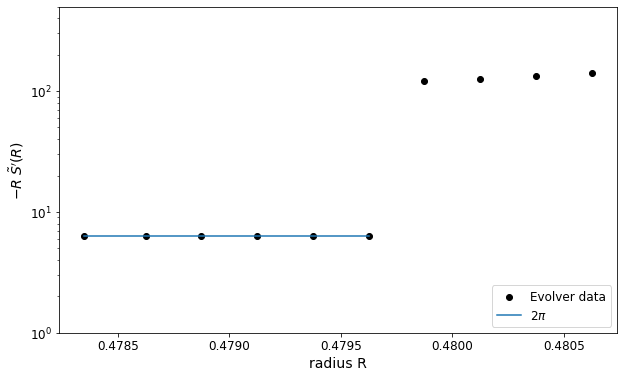}
     \end{subfigure}
    \caption{Left: The black data points show the area difference between the bridge surface and the cap surface at the same R. Right: The derivative of the EE is discontinuous across the transition from cap to bridge surfaces.}
    \label{fig:Ads5-4}
\end{figure}

For $R\lesssim 1/2$, the bridge configurations become important and have to be taken into account. The bridge surfaces have less symmetry and are harder to find, but they are ameanable to our numerical treatment. In this regime we have to compare the areas of the cap and bridge surfaces to determine which is dominant (their divergences are determined by near-boundary expansion and coincide, leaving a finite difference).
We find that close to a critical value $$R_{\rm bridge}\approx 0.48$$ the dominant minimal surface switches from the cap to the bridge configuration, just as in Fig. \ref{fig:topology_evolution}. The left panel of Fig. \ref{fig:Ads5-4} shows the difference in the area of the two configurations for varying $R$. At $R\sim0.478$, the bridge first appears as a locally minimal configuration, but the cap still has the globally minimal area. However, past $R\sim0.480$ the bridge configuration has the smaller area. The right panel shows the result of applying a derivative with respect to $R$. This shows a discontinuity; the slope of the EE changes across the phase transition.

For $R>1/2$, the entangling region develops creases and is no longer spherical. It only has a symmetry under rotation around the $x_1$ direction in (\ref{periodic_ads_metric}). It is then convenient to introduce a  coordinate system with polar coordinates $(\rho,\theta)$ on the $x_2,x_3$ plane: 
$$ds=\frac{1}{z^2}\left(-dt^2+dz^2+dx_1^2+d\rho^2+\rho^2d\theta^2\right).$$
Given the remaining rotational symmetry along $\theta$, the surface will be defined by a function $z(\rho,x_1)$ and the area can be minimized numerically, using an effective functional for a 2d surface in the 3-dimensional space $(\rho, x_1, z)$.

The sharp creases in the boundary of the entangling region for $R>1/2$ contribute $1/\epsilon$ terms \cite{Myers2012}. The general $\epsilon$ expansion for the full range of $R$ therefore is
\begin{equation}
    \text{Area}(\gamma_{\cal A}) =\frac{c_2}{\epsilon^2} + \frac{c_1}{\epsilon} + c_l\log\left(\frac{2R}{\epsilon}\right) + c_0 + \mathcal{O}(\epsilon^2).
    \label{AdS5_corners}
\end{equation}
The coefficient of the leading term is half the area of the entangling surface,
\begin{align}
c_2&=\frac{1}{2}\min\left(4\pi R^2,2\pi R \ell_1\right)\,.
\end{align}
The behavior for small and large $R$ is consistent with the discussion in section \ref{sec:across_dim}, but compared to the $\rm AdS_4$ example we do not have a maximum in between.
The remaining expansion coefficients in (\ref{AdS5_corners}) can be read off from (\ref{sphere exp}) for the cap surfaces. For the bridge surfaces the finite part is modified, as shown in Figure \ref{fig:Ads5-4} and Table \ref{tab:torus_bridge_coeff}.

\begin{table}[h!]
\hskip 10mm
\begin{subtable}{0.3\textwidth}
\centering
\begin{tabular}{ |c||c| }
 \hline
 $R$ & $\Delta A$\\
 \hline
  0.478 & 0.345\\
  0.479 & 0.178\\
  0.480 & -0.0405\\
  0.482 & -0.647\\
  0.485 & -2.14\\
  0.490 & -7.79\\
  0.495 & -27.2\\
  0.4975 & -55.5\\
 \hline
\end{tabular}
\caption{\label{tab:torus_bridge_coeff}}
\end{subtable}
\begin{subtable}{0.7\textwidth}
\centering
\begin{tabular}{ |c||c|c|c|c|c|  }
 \hline
 $R$ & $c_1$ & $c_l$ & $c_0$\\
 \hline
  0.52 & -1.30 & 0.008 & -1.39\\
  0.55 & -1.19 & 0.006 & -1.37\\
  0.6 & -1.06 & 0.005 & -1.33 \\
  0.8 & -0.757 & 0.009 & -1.24\\
  1 & -0.594 & 0.003 & -1.12\\
  1.5 & -0.389 & -0.009 & -0.896\\
  2 & -0.290 & -0.009 & -0.763\\
 \hline
\end{tabular}
\caption{\label{tab:torus_corner_coeff}}
\end{subtable}
\caption{Left: The (finite) area difference $\Delta A=A_{\rm bridge}-A_{\rm cap}$ between the bridge and cap surfaces for a range of $R$. Right: Entropy coefficients defined in Eq. (\ref{AdS5_corners}) for the corner phase.}
\end{table}

The numerical results for the corner regime $R>1/2$ are shown in Table \ref{tab:torus_corner_coeff}, with the $1/\epsilon$ terms from the creases clearly visible. The coefficient of the $1/\epsilon$ term tends to zero as $R$ increases, since the surface smoothes out. Meanwhile, $c_l$ changes from $-2\pi$ to roughly zero when the corners appear, and it continues to stay close to zero as $R$ increases further. The numerical data is consistent with $c_l$ actually being zero in the corner regime.

As the size of the entangling region $R$ is increased further beyond $\ell_1$, the creases smooth out. For $\ell_1\ll \ell_2$ this still happens well before the transition to the $\rm AdS_3\times T^2$ geometry. Once the singularities are smoothed out, the entangling region can be approximated as simply wrapping the shorter cycle of $T^2$ entirely. Since $R$ still is much smaller than the size of the longer cycle $\ell_2$ in the transition region to the  $\rm AdS_3\times T^2$ IR geometry, the EE changes smoothly through this transition region. Upon further increasing $R$, the surface eventually becomes comparable in size to $\ell_2$, and we expect another sequence of topology transitions of a similar form.
A difference is that the warp factor $g$ of the compact part of the geometry in (\ref{5D sugra metric main}) has become constant in this regime, and the $T^2$ has become part of the internal space. EE's associated with decompositions of the QFT degrees of freedom according to their representation in the internal space of the holographic dual were studied in flows across dimensions in \cite{Uhlemann:2021itz}. Here we have a hybrid situation, as the minimal surfaces at the transition $R\sim\ell_2$ still split both the CFT$_d$ geometry and the internal space.

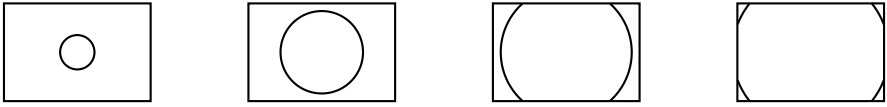
\begin{figure}
\centering
\begin{tikzpicture}[scale=1.3]
	\draw[thick] (0,0) rectangle +(1.5,1);
	\draw[thick] (0.75,0.5) circle (5pt);	
	\draw[thick] (2.5,0) rectangle +(1.5,1);
	\draw[thick] (3.25,0.5) circle (12pt);
	\draw[thick] (5,0) rectangle +(1.5,1);
	\draw[thick] [domain=-48:48] plot ({5.75+0.67*cos(\x)}, {0.5+0.67*sin(\x)});
	\draw[thick] [domain=-48:48] plot ({5.75-0.67*cos(\x)}, {0.5+0.67*sin(\x)});
	
	\draw[thick] (7.5,0) rectangle +(1.5,1);
	\draw[thick] [domain=21:39] plot ({8.25+0.8*cos(\x)}, {0.5+0.8*sin(\x)});
	\draw[thick] [domain=21:39] plot ({8.25-0.8*cos(\x)}, {0.5+0.8*sin(\x)});
	\draw[thick] [domain=21:39] plot ({8.25+0.8*cos(\x)}, {0.5-0.8*sin(\x)});
	\draw[thick] [domain=21:39] plot ({8.25-0.8*cos(\x)}, {0.5-0.8*sin(\x)});
\end{tikzpicture}
\caption{$T^2$ as rectangle with opposing edges identified, with a sequence of entangling regions of increasing size (showing sections at $x=0$, as described in the text). From left to right the cap phase, the bridge phase associated with the shorter cycle where the bridge extends across the horizontal edges, the bridge phase associated with the longer cycle where the bridge extends along the vertical edges, and the transition towards filling in the torus completely where the bridge fills in the corners.\label{fig:torus-transitions}}
\end{figure}

When the lengths of the cycles $\ell_1$ and $\ell_2$ are comparable the transitions associated with them are not as clearly separated. But we expect a similar picture. Say we have a CFT with spatial geometry $ds^2=dx^2+ds^2_{T^2}$, and we consider an entangling region of radius $R$ centered at $x=0$ and a reference point on $T^2$. As $R$ is increased, one first encounters a cap phase, then a sequence of bridge and corner phases associated with the shorter cycle, as described above, and then a second sequence of transitions associated with the longer cycle. Finally, we expect a third transition associated with filling the torus completely. This is illustrated in Fig.~\ref{fig:torus-transitions}. Once the entangling region includes almost all of the torus at $x=0$, as in Fig.~\ref{fig:torus-transitions} on the right, it may become beneficial for the RT surface to fill in the remaining part of the torus for some values of the AdS radial coordinate. This type of transition will be discussed in the next section for an $S^2$ compactification, where it can be exhibited clearly.

\subsection{4d  ${\cal N}=4$ SYM on $\mathbb{S}^2$} \label{N=4}

In this section we study a  compactification of ${\cal N}=4$ SYM on a 2-sphere, $\mathbb{S}^2$, and track the EE for a maximally spherical entangling region from the UV to the IR. The metric describing the RG flow holographically is (see appendix \ref{App:SugraSolutions} for details)
\begin{equation}\label{sphere metric main}
    ds^2=e^{2f(z)}(-dt^2+dx^2) + \frac{dz^2}{z^2 D(z)^2} + e^{2g(z)}(d\theta^2+\sin^2\theta d\phi^2)\;,
\end{equation}
where $\theta\in[0,\pi]$ and $\phi\in[0,2\pi)$. 
Plots of an example solution are shown in Fig.~\ref{fig:sphere_fgd}.
The 4d geometry on which the UV CFT is defined takes the form 
\begin{equation}
    ds^2=-dt^2+dx^2 + d\theta^2+\sin^2\!\theta\, d\phi^2\;.
\end{equation}
We will study EE's associated with spherical regions centered at $x=\theta=0$ on a constant time slice. More precisely, the regions are defined by $x^2+\theta^2\leq R^2$.
The metric and entangling regions are symmetric under rotations along $\phi$, so that the corresponding minimal surfaces can be defined by a function $z(x,\theta)$. The universal term in the UV will give the $a$ central charge, and we can follow the evolution of the EE all the way into the IR.

\begin{figure}
    \centering
    \includegraphics[width=0.3\textwidth]{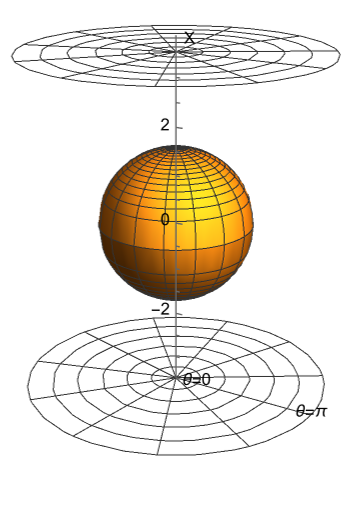}
    \caption{Maximally spherical entangling region with $R=1.6.$}
    \label{fig:sphere on sphere}
\end{figure}

We can graphically show the entanglement regions using plots covering the spatial directions $(\theta,\phi,x)$. An example is shown in Fig.~\ref{fig:sphere on sphere} with a spherical entangling region in orange. In this plot, the $(\theta,\phi)$ part is displayed as a disc with $\theta$ increasing radially and $\phi$ in the angular direction. The non-periodic coordinate $x$ points upward. This representation captures the geometry near the $\theta=0$ pole at the origin of the disc well, while the entire $\theta=\pi$ boundary of the $(\theta,\phi)$ disc at each fixed $x$ represents a single point.

\begin{figure}
     \centering
     \begin{subfigure}[h!]{0.30\textwidth}
         \centering
         \includegraphics[width=0.93\textwidth]{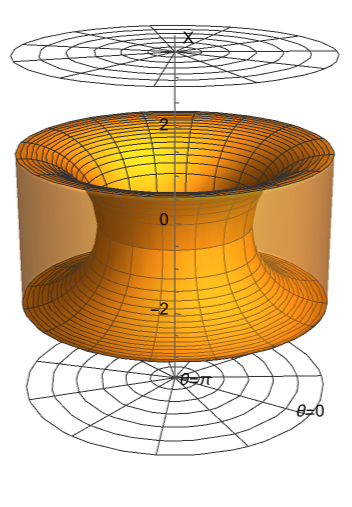}
         \caption{R=1.6}
     \end{subfigure}
     \hspace{10 pt}
     \begin{subfigure}[h!]{0.30\textwidth}
         \centering
         \includegraphics[width=0.93\textwidth]{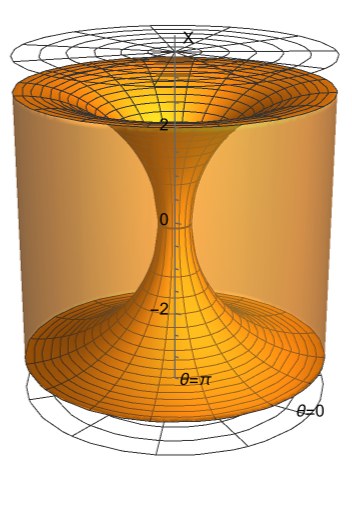}
         \caption{R=2.8}
     \end{subfigure}
     \hspace{10 pt}
     \begin{subfigure}[h!]{0.30\textwidth}
         \centering
         \includegraphics[width=0.93\textwidth]{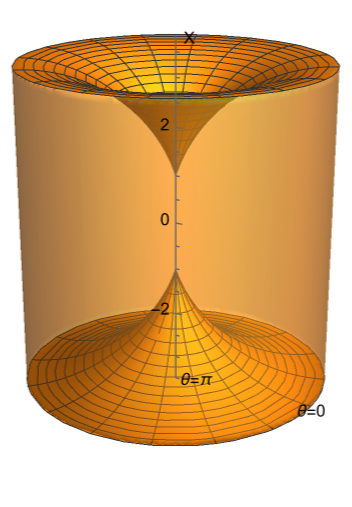}
         \caption{R=3.3}
     \end{subfigure}
     \vskip\baselineskip
     \begin{subfigure}[h!]{0.3\textwidth}
         \centering
         \includegraphics[width=\textwidth]{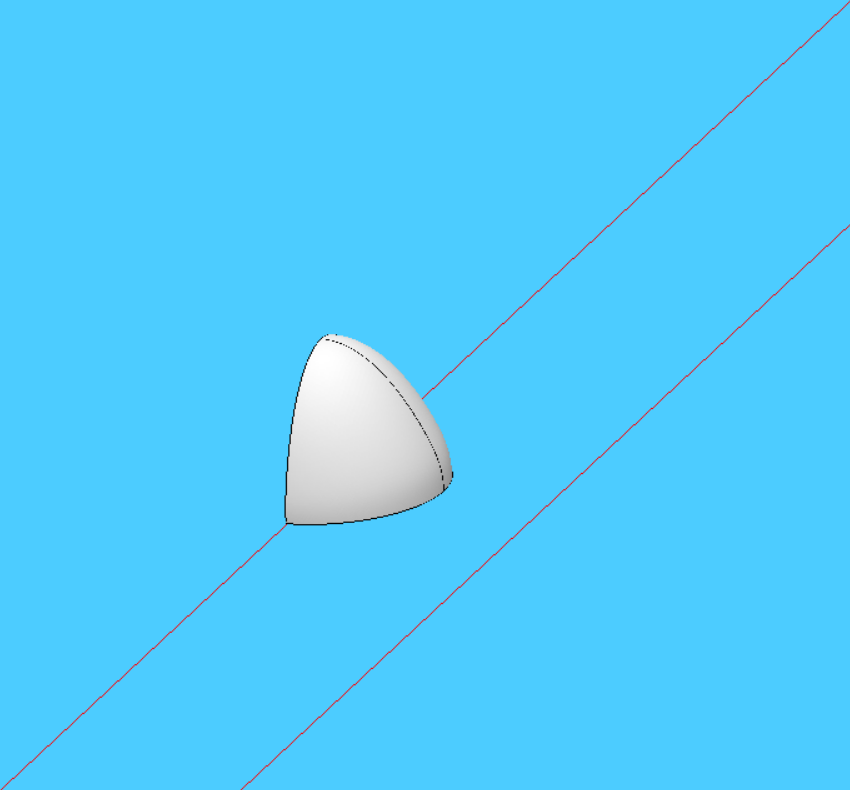}
     \end{subfigure}
     \hspace{10 pt}
     \begin{subfigure}[h!]{0.3\textwidth}
         \centering
         \includegraphics[width=\textwidth]{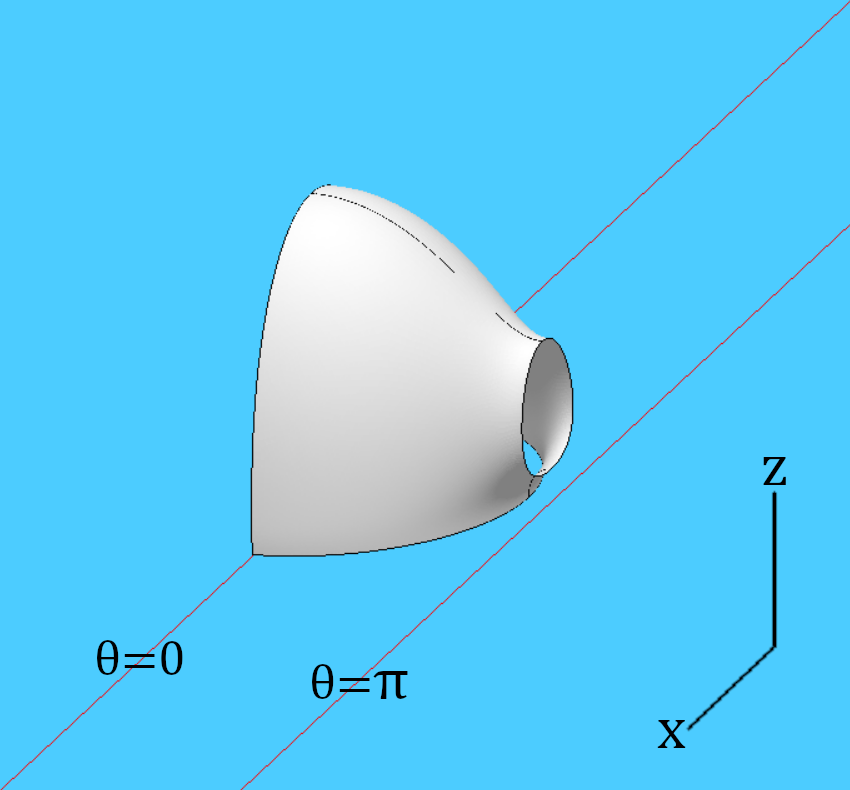}
     \end{subfigure}
     \hspace{10 pt}
     \begin{subfigure}[h!]{0.3\textwidth}
         \centering
         \includegraphics[width=\textwidth]{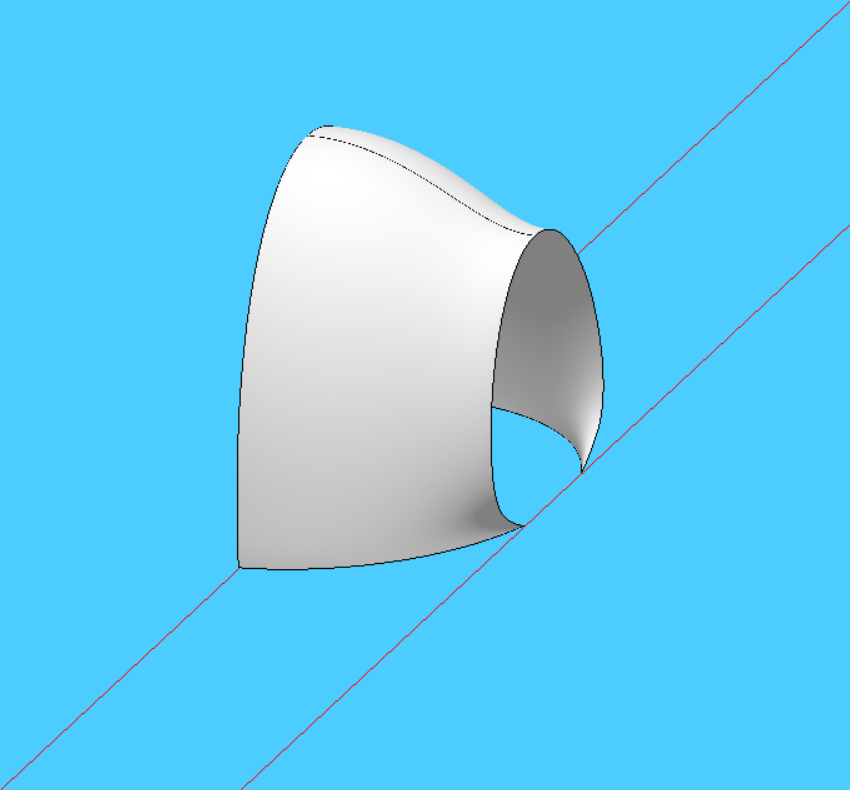}
     \end{subfigure}
        \caption{Top: Entangling regions in ${\cal N}=4$ SYM on $S^2$ for increasing values of $R$. This is the sphere analog of the process in Fig. \ref{fig:boundary_evolution}. Bottom: Minimal surfaces of the functional (\ref{sphere area integral}). The direction $\phi$ is omitted, since the surface possesses axial symmetry.}
        \label{fig:phase_trans_sphere}
\end{figure}

Once the entangling region grows such that it nearly wraps the $S^2$ at $x=0$, it will be more convenient to instead choose $\theta=\pi$ as the central vertical line and have $\theta$ increase inwards. This is shown in Fig.~\ref{fig:phase_trans_sphere}. The $\theta=\pi$ pole is now at the center of the disc, while the entire boundary corresponds to the $\theta=0$ pole.
Fig.~\ref{fig:phase_trans_sphere}(a) shows the same region as in Fig. \ref{fig:sphere on sphere} in this new representation. The interior of the entangling region (shown in semi-transparent orange) still extends over the same range of coordinates as in the previous figure. At $x=0$ for example, the region covers $\theta\in[0,R]$ and $\phi\in[0,2\pi)$ in both figures. However, the behavior at $\theta=\pi$ is now more transparent. As the size of the region increases further in (b) and (c), the region eventually covers all of $(\theta,\phi)$ for a certain range of $x$ in (c). Corners appear in (c), this time with the geometry of a cone. These lead to $\log^2(1/\epsilon)$ terms in the EE \cite{Myers2012} (as compared to the $1/\epsilon$ terms which appeared for the torus). As $R$ increases further, the region continues to expand in the $x$ directions and the corners smooth out.

The bottom row of Fig.~\ref{fig:phase_trans_sphere} shows the minimal surfaces associated with the respective regions in the top row. 
Due to the symmetry under rotations along $\phi$, the minimal surfaces can be determined using an effectively 2-dimensional functional (\ref{sphere area integral}) in $(x,\theta,z)$ space. In the figures, the two red lines mark the limits for $\theta\in[0,\pi)$, and $z$ points upwards. With increasing $R$, the minimal surface evolves through similar stages as for the $S^1$ model in Sec.~\ref{sec:across_dim} before. Compared to the torus, there are no 1-cycles and the only transition is associated with the RT surface starting to wrap the entire $S^2$ at some point. The surface begins in the cap configuration for small $R$ before undergoing a phase transition in (b). Note that the new protrusion in (b) is not visible in the shape of the entangling region above, since it only exists at some range of $z$ greater than zero. Before this transition, the surface does not extend into the range $\theta \in [R,\pi)$. After the transition it does for a certain range of $x$ and $z$. 

\begin{figure}
    \centering
    \includegraphics[width=0.86\textwidth]{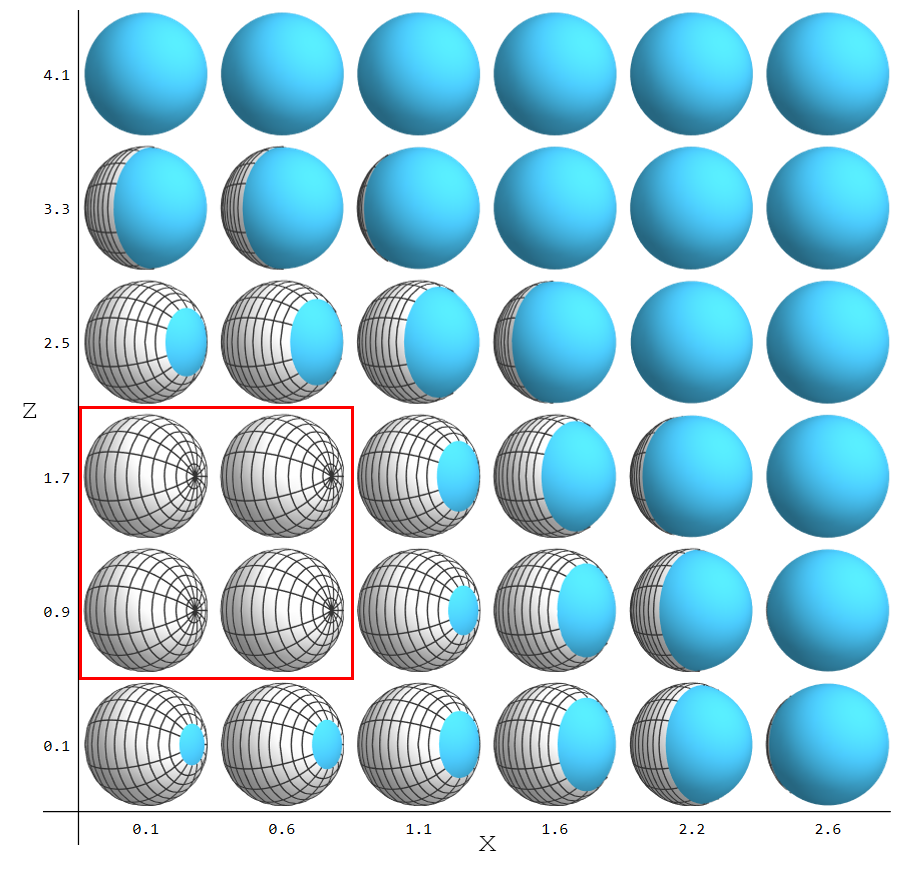}
    \caption{Array of spheres representing the spatial part of the geometry (\ref{sphere metric main}) by factoring it into a $(\theta,\phi)$ sphere at each position $(x,z)$. The bottom line is the conformal boundary with the entangling region. The surface in fig. \ref{fig:phase_trans_sphere}(b) is shown in gray. The red square highlights the position of the bridge.}
    \label{fig:sphere_array}
\end{figure}

Another visualization of the surface in fig.~\ref{fig:phase_trans_sphere}(b) is shown in Fig. \ref{fig:sphere_array}. Here, the $x$ and $z$ coordinates run horizontally and vertically, respectively, and the sphere at each position on the grid represents the corresponding ($\theta,\phi$). The color scheme remains the same, with the RT surface shown in gray on a blue background. The effect of the aforementioned phase transition is visible in the red square, where the surface now wraps the compact space completely. The new topology may again be aptly called a ``bridge'' configuration, since it adds a connection across the South pole of the spheres for some range of $z$.

As before, we may extract the coefficients of the $\epsilon$ expansion
\begin{equation}
    \text{Area}(\gamma_{\cal A}) =\frac{c_2}{\epsilon^2} + c_{l2}\log^2\left(\frac{2\pi}{\epsilon}\right) + c_l\log\left(\frac{2R}{\epsilon}\right) + c_0 + \mathcal{O}(\epsilon^2).
    \label{S2_EE_expansion}
\end{equation}
Here, the circumference of the $(\theta,\phi)$ sphere is chosen as the length scale in the $\log^2$ term. This choice is conventional and the ambiguity has the consequence that $c_l$ is no longer universal.

The $c_2$ and $c_{l2}$ coefficients can be computed analytically and used in the fits for the remaining terms. As before, $c_2$ is half of the boundary area of the entangling region. Using the parametrization $(x,\theta,\phi) = (R\cos{\alpha}, R\sin{\alpha},\phi)$ for the boundary,
\begin{equation}
    c_2 = 2\pi\int_0^{\alpha_0} R \sin{(R\sin{\alpha})}\; d\alpha
\end{equation}
with 
\begin{equation}
    \alpha_0 = 
    \begin{cases}
        \frac{\pi}{2}, & \text{for } R<\pi\\
        \arcsin\left(\frac{\pi}{R} \right), & \text{for } R>\pi.
    \end{cases}
\end{equation}
The behavior of $c_2$ as function of $R$ is in line with the discussion in section \ref{sec:across_dim}: for small $R$ it grows as $2\pi R^2$ (half the area of a spherical entangling region in the 4-dimensional UV geometry), then goes through a maximum and ultimately asymptotes to $4\pi$. At $R=\pi$, $c_2$ has a continuous first derivative.
Meanwhile, $c_{l2}$ is given by (\ref{corner_log2}) for $R>\pi$ and is zero otherwise. These values are listed for a range of $R$ in Table \ref{tab:entropy_coeff_S2} along with numerical results for the remaining coefficients in (\ref{S2_EE_expansion}).
We note that these EE's are obtained from minimal surfaces in a full supergravity flow solution, which was obtained numerically. As a result, we do not have a closed form expression for the EE in the cap phase (unlike for the torus in the thin limit). However, for small $R$ the values of $c_l$ and $c_0$ approach those in (\ref{sphere exp}).

\begin{table}
\begin{center}
		\begin{tabular}{||c|c|c|c||c|c|c|c||c|c|c|c||}
			\hline
			\multicolumn{4}{||c||}{Cap/UV}  & 			\multicolumn{4}{|c||}{Bridge} &
			\multicolumn{4}{|c||}{Corners} \\
			\hline
			$R$ & $c_{l2}$ & $c_l$ & $c_0$ & $R$ & $c_{l2}$ & $c_l$ & $c_0$ & $R$ & $c_{l2}$ & $c_l$ & $c_0$\\
			\hline
			0.1 & 0 & -6.49 & 3.28 & 2.72 & 0 & -52.4 & 139 & 3.2 & -3.98 & -23.5 & 105\\
			0.2 & 0 & -7.02 & 3.61 & 2.9 & 0 & -54.2 & 149 & 3.3 & -2.33 & -33.7 & 127\\
			0.5 & 0 & -9.94 & 5.27 & 			 3.0 & 0 & -56.8 & 160 & 3.5 & -1.44 & -37.3 & 136\\
			1.0 & 0 & -19.4 & 19.6 & 			 3.1 & 0 & -72.3 & 228 & 4.0 & -0.783 & -38.3 & 145 \\
   \cline{5-8}
			2.0 & 0 & -43.6 & 88.3 &\multicolumn{4}{c||}{}& 6.0 & -0.253 & -38.1 & 166\\
			2.7 & 0 & -52.2 & 137 &\multicolumn{4}{c||}{}& 			 10.0 & -0.0816 & -37.8 & 192\\
			\cline{1-4}\cline{9-12}
		\end{tabular}
\end{center}
\caption{Entropy coefficients defined in Eq. (\ref{S2_EE_expansion}). The range of $R$ values is divided into three intervals corresponding to the cap, bridge, and corner geometries.}
\label{tab:entropy_coeff_S2}
\end{table}

\begin{figure}
     \centering
     \begin{subfigure}[h!]{0.49\textwidth}
         \centering
         \includegraphics[width=\textwidth]{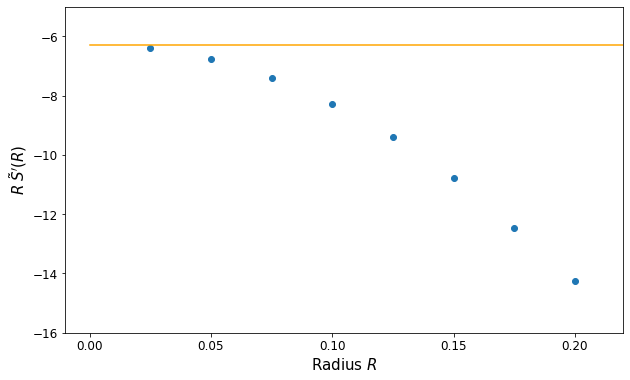}
         \caption{}
     \end{subfigure}
     \hspace{0 pt}
     \begin{subfigure}[h!]{0.49\textwidth}
         \centering
         \includegraphics[width=\textwidth]{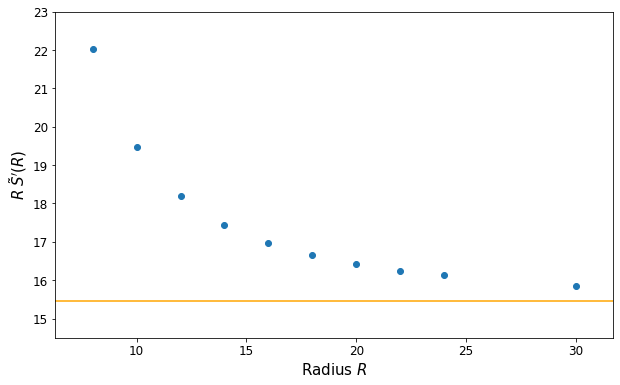}
         \caption{}
     \end{subfigure}
     \begin{subfigure}[h!]{0.49\textwidth}
         \centering
         \includegraphics[width=\textwidth]{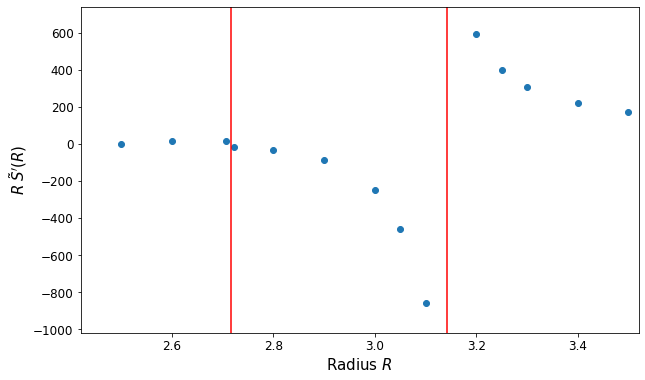}
         \caption{}
     \end{subfigure}
        \caption{Evolution of $R\tilde{S}'(R)$ for ${\cal N}=4$ SYM on $S^2$. Graphs (a) and (b) show the behavior for small and large entangling radius $R$, respectively. The interpolating function matches the expected values in the UV and IR (orange segments). Graph (c) shows the behavior in an intermediate range where the function has discontinuities. The red vertical lines mark out the appearance of the bridge at $R\approx 2.72$ and the corners at $R=\pi.$}
        \label{fig:RSprime}
\end{figure}

Since the entangling region becomes spherical in the UV and wraps the compact directions uniformly in the IR limit, the EE encodes the central charges at the fixed points of the flow. 
Denoting by $\tilde{S}$ the EE with the $1/\epsilon^2$ divergence subtracted, the quantity $R\tilde{S}'(R)$ extracts the logarithmic term in both the UV and IR limits. Although intermediate entangling regions possess a $\log^2{\epsilon}$ term, this divergence will decrease with increasing $R$. 
Figure \ref{fig:RSprime} shows the evolution of $R\tilde{S}'(R)$ for both small and large entangling regions. The orange lines label the central charges
\begin{equation}
    c_{UV} = -2\pi, \hspace{1 in} c_{IR} = \frac{8\pi e^{2g_0}}{D_0},
\end{equation}
where $g_0 = \lim_{z\to\infty} g(z)$ and $D_0 = \lim_{z\to\infty} D(z)$. The numerical data agrees with both limits.
The third plot in Figure \ref{fig:RSprime} shows the discontinuity in the derivative of the EE at the transition to the bridge phase (similar to Fig.~\ref{fig:Ads5-4}) and the behavior at the transition to the corner phase. The growth of the derivative just before the onset of corners may indicate that the bridges become sharp when approaching the corner regime. We note  that the $1/\epsilon^2$ divergence, which was subtracted by hand in the definition of $\tilde S$, also has non-trivial $R$-dependence; the plots simply illustrate the behavior of the remaining terms.

\subsection{Higher genus surfaces}\label{sec:higher-genus}

After discussing the genus zero and one cases in detail, we can formulate our expectations for hyperbolic Riemann surfaces of higher genus. These may be seen as the most basic case, in the sense that supersymmetric compactifications on hyperbolic surfaces can typically be realized using universal twists based on the R-symmetry alone. 

As a closed form example of a flow solution, we can take the BPS magnetic black string solution of 5d ${\cal N}=2$ gauged supergravity presented in \cite{Klemm:2000nj}. The metric takes the form
\begin{eqnarray}
    ds^2 &=& \frac{L^2}{z^2}\left(1-\frac{z^2}{3L^2}\right)^{3/2}(-dt^2+dr^2) + \frac{L^2dz^2}{z^2 \left(1-\frac{z^2}{3L^2}\right)^2} + \frac{L^2}{z^2}d\Sigma^2.
\end{eqnarray}
The 5d ${\cal N}=2$ supergravity background also contains magnetic fluxes along the Riemann surface $\Sigma$ and non-trivial scalars. 
The metric becomes AdS$_5$ with foliation $\mathbb{R}^{1,1}\times \Sigma$ for $z\to 0$ in the UV and AdS$_3\times \Sigma$ for $z\to \sqrt{3}L$ in the IR.
As starting point of our discussion we again consider a small spherical region with $R\ll \ell$ probing the UV physcis. The EE takes the form
\begin{equation}
    S_{EE}(R,\epsilon) = \frac{2\pi R^2}{\epsilon^2} - 2\pi\log\left(\frac{R}{\epsilon}\right) +\mathcal{O}(\epsilon^0)\,,
\end{equation}
and the coefficient of the log term gives the UV central charge. In the opposite limit, for large entangling regions, any corner contributions have smoothed out and the entangling region fully wraps the compact space, such that $\Sigma$ affects the EE only through an overall factor of its volume. The expansion of the EE then is instead
\begin{equation}
    S_{EE}(R,L,\epsilon) = \frac{\ell^2}{\epsilon^2}+\frac{2\ell^2}{3L^2}\log\left(\frac{L}{\epsilon}\right)-\frac{2\ell^2}{3L R}+\frac{4 \ell^2}{9 L^2}\log\left(\frac{R}{L}\right)+\cdots
\end{equation}
The $\log \epsilon$ coefficient now gives a linear combination of the UV central charges, as demonstrated in \cite{GonzalezLezcano:2022mcd}.
The IR central charge appears as coefficient of the $\log R$ term.  Along the flow, where $R$ and $\ell$ are comparable, such expansions do not apply, but we may still define interpolating functions by applying appropriate derivative operations with respect to $R$ to $S_{EE}$, which extract the appropriate $\log$ terms at the fixed points.

In analogy to the realization of the torus as quotient $\CC/(\ZZ+\tau\ZZ)$, hyperbolic Riemann surfaces can be realized as quotients of the hyperbolic plane $\mathds{H}_2$, as $\Sigma_{g>1}=\mathds{H}_2/\Gamma$, with $\Gamma$ a Fuchsian group. We therefore expect a similar sequence of phases in the EE for a region of increasing size: Initially, for small regions, the fact that the QFT is defined on $\RR^{1,1}\times \mathds{H}_2/\Gamma$ instead of $\RR^{1,1}\times \mathds{H}_2$ does not lead to qualitative changes in the EE. The RT surface is a cap, similar to the torus case, though the curvature of $\mathds{H}_2$ in principle affects both the definition of a (generalized) spherical region following (\ref{eq:region-def}) and the associated RT surfaces.
As the size of the region is increased, it will eventually detect the identifications resulting from the quotient. One first encounters a regime where the RT surfaces associated with the regions in neighboring fundamental domains almost touch. Following the arguments of sec.~\ref{sec:bridge_argument}, this leads to a phase with bridges.
As the size of the region is increased further, these bridges turn into corner type singularities in the entangling region, leading to new terms in the EE, whose impact decays as the size of the region is increased further and the singularities smooth out.

For higher genus surfaces we expect one such sequence of transitions through bridge and corner phases for each 1-cycle, with the order of the transitions associated with different cycles and the associated scales of the entangling regions depending on the geometry and topology of the compact space and the placement of the entangling region. Eventually, there should again also be a transition associated with filling the entire surface, analogously to the transition for $S^2$. We will not enter into a quantitative discussion here, but it would certainly be interesting to further study this connection between the EE and the topology of the compact manifold.

\section{Conclusions}\label{Sec:Conclusions}

We tracked the evolution of the entanglement entropy in RG flows across dimensions as a small spherical region probing the UV physics grows in size to increasingly probe the IR phyics. The starting point in the UV is the region whose EE would naturally lead to a $c$-function for flows within the higher-dimensional theory. The end point in the IR is a region wrapping the compact space entirely while being spherical in the remaining non-compact directions; this is the region whose EE would lead to a c-function in the lower-dimensional theory. 

Along the evolution we identified universal sequences of transitions through what we called cap, bridge and corner phases. The cap phases refer to the regime where the bulk RT surface does not detect the partial compactification yet. The bridge phase refers to regimes where the RT surface in the bulk wraps a part of the compact space which is larger than the entangling region. When the compact space is represented as a quotient, this can be understood as the RT surface forming ``bridges" to copies of itself in adjacent fundamental domains. Finally, the corner phase arises from entangling regions which are large enough to develop corners or creases as a result of the (partially) compact QFT geometry. When the compact space is represented as a quotient this can be understood as the entangling region intersecting copies of itself in adjacent fundamental domains.

These sequences of transitions are naturally associated with cycles in the compact part of the QFT geometry. For compactifications of 4d $\mathcal N=4$ SYM on 2d Riemann surfaces we identified transitions associated with 1-cylces (demonstrated for torus compactifications in sec.~\ref{Subsect:ThinTorus}) and transitions associated with the compact manifold itself (demonstrated for $S^2$ compactifications in sec.~\ref{N=4}). Both types of transitions come with bridge and corner phases. 

In the transition to the bridge phase, the EE changes continuously (though with discontinuous derivative, as shown in fig.~\ref{fig:Ads5-4}). The transition to the corner/crease phase, on the other hand, leads to new divergences. The corners are sharpest and the resulting singularities in the EE are strongest immediately after their onset (when the size of the entangling region just barely exceeds the length of the cycle). They smooth out afterwards and eventually disappear as the size of the entangling region becomes large. These divergences are easy to track and their contribution to the EE known analytically, so they give an entanglement handle on the geometry and topology of the compact part of the QFT geometry.

The results discussed above are also interesting in the context of interpolating functions connecting the UV and IR central charges in flows across dimensions. 
Candidate functions  were studied e.g.\ in \cite{Macpherson:2014eza,GonzalezLezcano:2022mcd,Bea:2015fja,Chu:2019uoh,Legramandi:2021aqv,Merrikin:2022yho}.
The evolution of entangling regions we considered here is perhaps the most natural choice for the purpose of defining functions that connect the UV and IR central charges. 
As explained in \cite{GonzalezLezcano:2022mcd}, one may not expect functions interpolating between central charges in different dimensions to be monotonic, since central charges in different dimensions count different notions of degrees of freedom.
However, interpolating functions based on the sequences of entangling regions considered here capture divergences arising along the flow from corners and creases and  encode the geometry and topology of the compact space in their scale dependence. This makes them an interesting subject of study.

As avenues for future research it would be interesting to further flesh out the connection between the EE and the features of the compact space, for example by studying higher-dimensional compact spaces, higher-genus Riemann surfaces (expanding on section \ref{sec:higher-genus}). Further investigating interpolating functions, and developing a  field theory understanding are other interesting open directions.

\section*{Acknowledgments}
We are grateful to  Alfredo Gonz\'alez Lezcano, Sebastian Grieninger,  Junho Hong, Imtak Jeon, James T. Liu, Augniva Ray and  Erik Tonni for discussions. We are particularly thankful to Horacio Casini for various clarifying discussions. This work is supported in part by the U.S. Department of Energy under grant DE-SC0007859. ED was supported in part by a Leinweber Graduate Summer  Fellowship. LPZ is grateful to APCTP for hospitality as a Senior Advisory Group member, he also  acknowledges support from an IBM Einstein Fellowship at the Institute for Advanced Study.

\appendix

\section{Surface Evolver: technical details}\label{App:SurfaceEvolver}
In the main part we use the \textit{Surface Evolver} \cite{Brakke,Brakke1992}, a program that constructs approximations to minimal surfaces for given boundary conditions using triangulations. In this appendix we demonstrate its efficacy by explaining convergence properties, the modifications we need to implement, and by confronting the results  with analytic expectations.

\subsection{Implementing area functionals} \label{sec:area functional}
The Surface Evolver computes the area of a triangular mesh by integrating a user-defined functional over each triangle and adding the results. The Evolver supports non-Euclidean metrics and can compute areas directly. However, for surfaces in more than three spatial dimensions this method becomes time consuming. Instead, we can consider effective surfaces in three dimensions, and specify a functional to integrate. In all cases, the surfaces will be considered to exist in Euclidean space. The metric capabilities of the Evolver can actually be repurposed in a way that will be described later.

Consider a codimension-1 surface in $n$-dimensional space. Suppose that the surface is symmetric along the directions $x_4, x_5,\ldots x_n$ so that it can be defined by an equation $x_3 = x_3(x_1,x_2)$. The area is
\begin{equation}
    \text{Area} = \int_R \sqrt{\det g_{ind}}\; dx_1 dx_2,
\end{equation}
where $R$ defines the region in the $x_1,x_2$ plane and $g_{ind}$ is the induced metric on the surface. Consider a general diagonal metric
\begin{equation}
    ds^2 = Adx_1^2 + Bdx_2^2 + Cdx_3^2 + \sum_{i>3} D_i dx_i^2,
\end{equation}
where $A$, $B$, $C$ and the $D_i$ are functions of all coordinates $x_1,x_2,\ldots x_n$. Utilizing the equation for $x_3$ we have
\begin{equation}
    dx_3 = \frac{\partial x_3}{\partial x_1}dx_1 + \frac{\partial x_3}{\partial x_2}dx_2,
\end{equation}
so
\begin{equation}
    ds^2 = \left(A + C\left(\frac{\partial x_3}{\partial x_1}\right)^2\right)dx_1^2 + \left(B + C\left(\frac{\partial x_3}{\partial x_2}\right)^2\right)dx_2^2 + 2C\frac{\partial x_3}{\partial x_1}\frac{\partial x_3}{\partial x_2}dx_1 dx_2 + \cdots.
\end{equation}
The determinant is
\begin{equation}
    \det{g_{ind}} = \left(\prod_{i>3} D_i\right)\left(AB + BC\left(\frac{\partial x_3}{\partial x_1}\right)^2 + AC\left(\frac{\partial x_3}{\partial x_2}\right)^2\right),
\end{equation}
and the area is
\begin{equation}\label{general area integral}
    \text{Area} = \bigintss_R D\sqrt{AB + BC\left(\frac{\partial x_3}{\partial x_1}\right)^2 + AC\left(\frac{\partial x_3}{\partial x_2}\right)^2} dx_1 dx_2.
\end{equation}
Here $D \equiv \sqrt{\prod_{i>3} D_i}$. There are multiple methods for implementing a functional in the Evolver, two of which were used for this manuscript.

\textbf{1) Integral over Euclidean area}
If $A=B=C$, as is the case in pure AdS, then we can pull all metric functions in (\ref{general area integral}) out of the square root and write the expression as the integral of a single function over the Euclidean area element $dA$:

\begin{equation} \label{int over Euc}
    \text{Area} = \bigintss_R f(x_1,\ldots,x_n) \sqrt{1 + \left(\frac{\partial x_3}{\partial x_1}\right)^2 + \left(\frac{\partial x_3}{\partial x_2}\right)^2} dx_1 dx_2 = \int_R f(x_1,\ldots,x_n) da.
\end{equation}
In the Evolver, the surface can therefore be treated as existing in Euclidean space, and the function $f$ can be defined by the user.

\textbf{2) Integral over a function dependent on surface normal}
For general metric functions $A, B$ and $C$ we must utilize a different method. The Evolver allows for integrals over functions that depend on both position and surface normal. For any triangular facet, the normal is defined by $\bf{n}=\bf{u}\times\bf{v}$, where $\bf{u}$ and $\bf{v}$ represent two edges of the triangle. With this method, one defines a function $f(\bf{x},\bf{n})$ such that the desired integral over the full surface is approximated by

\begin{equation} \label{facet_sum}
    \frac{1}{2}\sum_{F_i\in \text{Facets}} f(\mathbf{x},\mathbf{n}).
\end{equation}
With this convention, $f(\bf{x},\bf{n}) = |\bf{n}|$ gives the usual Euclidean area. We must then find a way to reproduce the square root term in (\ref{general area integral}) using the given definition for surface normal.
If we consider a small facet of the effective surface in $x_1,x_2,x_3$ space, its area is

\begin{equation}
    \text{Area}(\text{facet}) =\frac{1}{2} D \left|\sqrt{\det g}\epsilon_{ijk} u^j v^k\right|,
\end{equation}
where $g$ is the metric
\begin{equation}
    ds^2 = Adx_1^2 + Bdx_2^2 + Cdx_3^2,
\end{equation}
and the expression inside the norm is the generalization of the cross product to curved manifolds. But $n_i = \epsilon_{ijk} u^j v^k$ is just the Euclidean normal, so
\begin{align} \label{int surface normal}
    \text{Area}(\text{facet}) &= \frac{1}{2} D \left|\sqrt{\det g}\;n_i\right| = \frac{1}{2} D ABC\left|n_i\right| \nonumber\\
     &= \frac{1}{2} D ABC\sqrt{\frac{1}{A}n_1^2+\frac{1}{B}n_2^2+\frac{1}{C}n_3^2} \\
     &= \frac{1}{2} D \sqrt{BCn_1^2+ACn_2^2+ABn_3^2} \nonumber.
\end{align}
Comparing this expression with (\ref{facet_sum}) immediately determines $f(\mathbf{x},\mathbf{n})$.

\bigskip
{\bf The string model:}
The Surface Evolver also supports a lower-dimensional \textit{string model}, which minimizes a length functional on a curve by approximating it with line segments. This model is useful for highly symmetric entangling regions where the minimal surface is a function of only one coordinate. In the case where $x_3 = x_3(x_1)$, equation (\ref{int over Euc}) becomes 

\begin{equation}
    \text{Area} = \bigintss_R f(x_1,\ldots,x_n) \sqrt{1 + \left(\frac{\partial x_3}{\partial x_1}\right)^2} dx_1 = \int_R f(x_1,\ldots,x_n) dl.
\end{equation}
Thus the integral can be taken along the Euclidean length of a curve in ($x_1,x_3)$ space. Equation (\ref{int surface normal}) can be similarly simplified, although with a small modification since the Evolver uses a tangent vector $\mathbf{t}=(t_1, t_2)$ in the string model rather than a normal vector.

\bigskip
{\bf Further comments:}
For the surfaces considered in this manuscript, a small value of $\epsilon$ on the order of 0.01 times the length scale of the entangling region was chosen, and the boundary geometry was then enforced at $z=\epsilon$ rather than $z=0$. A rough initial triangulation was constructed based on the expected shape of the surface, and then input and refined in the Surface Evolver. However, using the usual refinement procedure in which each edge is divided exactly at the midpoint was not the most efficient method. For surfaces in AdS, the portion close to the boundary accounts for the largest part of the area. Using a variable triangle size, with smaller triangles near the boundary produces a more efficient triangulation. In the Evolver, the metric for the ambient space determines where the midpoint for edge subdivision should lie. To have more control over subdivision, a metric was implemented solely for this purpose without affecting the area functional. Usually, a conformal metric with factor $1/z$ or $1/z^2$ was chosen to bring the subdivision midpoints closer to the boundary. Sometimes, more complicated metrics were designed with additional dependence on other coordinates to create areas with denser triangulation.

\subsection{Results for AdS$_4$ and AdS$_5$}\label{app:AdS4}
\begin{figure}
     \centering
     \begin{subfigure}[h!]{0.38\textwidth}
         \centering
         \includegraphics[width=\textwidth]{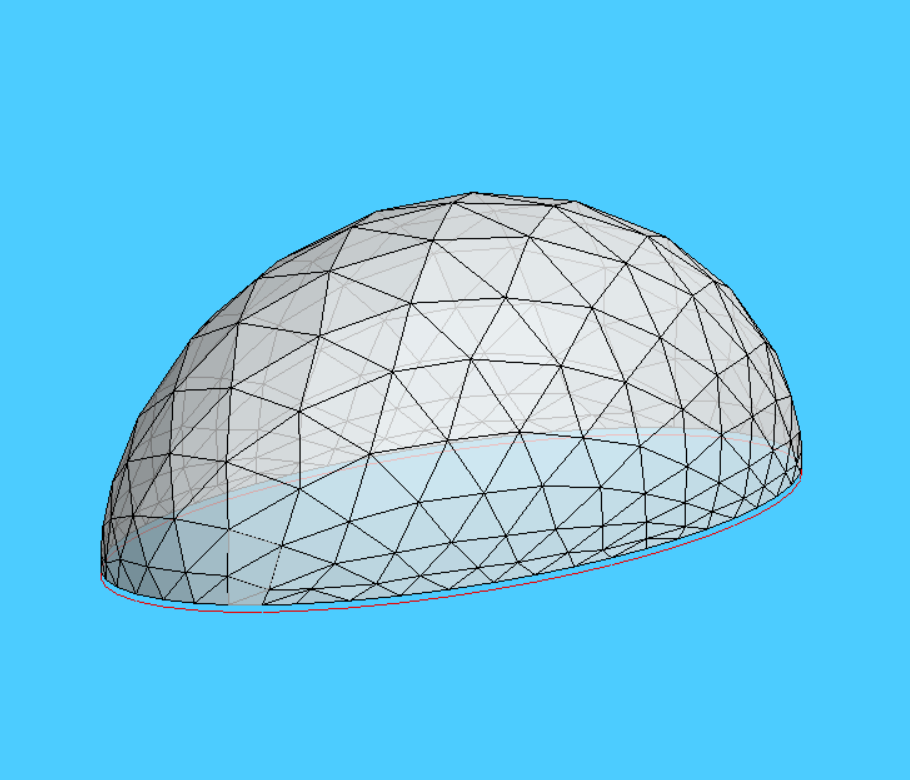}
     \end{subfigure}
     \qquad
     \begin{subfigure}[h!]{0.38\textwidth}
         \centering
         \includegraphics[width=\textwidth]{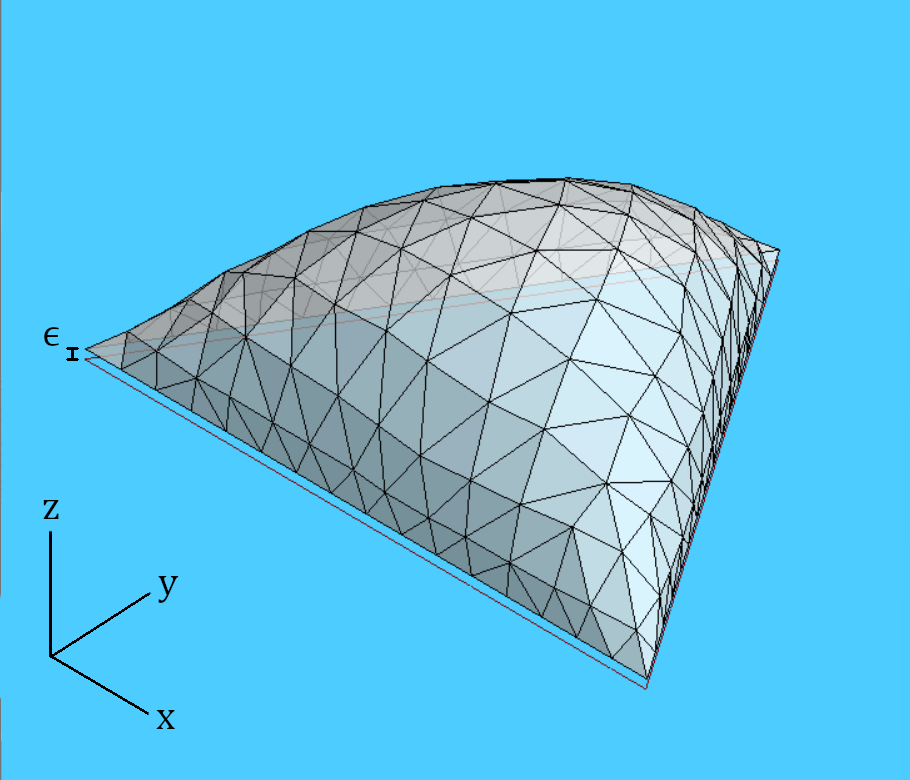}
     \end{subfigure}
        \caption{Approximate minimal surfaces for an ellipse and triangle in CFT$_3$. The elliptical boundary is defined by $x^2/4+y^2 = 1$ and the triangle has side length $\sqrt{3}$. In both figures, the holographic direction $z$ points upward. The boundaries of the entangling regions are shown in red, and the surfaces are cut off at separation $\epsilon=0.03$ above them. The triangulation on these surfaces is purposefully rough to demonstrate the process used by the Evolver.}
        \label{fig:CFT3_examples}
\end{figure}

Minimal surfaces in AdS$_4$ associated with regions in a CFT$_3$ are shown in the examples in fig. \ref{fig:CFT3_examples}. Surfaces in AdS$_4$ were already studied with the Evolver in \cite{Fonda2015}, and in this subsection we briefly reproduce one result as calibration.

Consider a rhombus of side length 1 with vertex angle $\alpha$. The area of the minimal surface is
\begin{equation}\label{rhombus_expansion}
    \text{Area}(\gamma_{\cal A}) = \frac{4}{\epsilon} + c_l\log\left(\frac{1}{\epsilon}\right) + c_0 + \mathcal{O}(\epsilon),
\end{equation}
where the coefficient of the $1/\epsilon$ term comes from the perimeter of the boundary, and the logarithmic term arises only from the corners.

\begin{figure}[h!]
    \centering
    \includegraphics[width=0.55\textwidth]{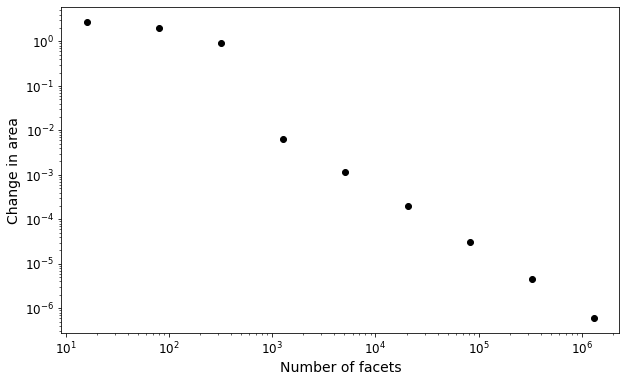}
    \caption{{\bf Convergence} Change in area after successive refinements of the minimal surface for a rhombus. Here the side length is 1, vertex angle is $\pi/3$, and cutoff is $\epsilon=0.03$. The large change at around 1300 facets is do to a switch from the linear to quadratic surface interpolation model. After this point the graph is nearly linear, with the area modifications growing smaller by more than an order of magnitude at each refinement. Assuming such behavior continues, convergence is guaranteed.}
    \label{fig:rhombus_accuracy}
\end{figure}

In the Evolver, a pyramid composed of four triangles above the rhombus can be chosen as the initial trial surface. However, since the rhombus has two reflection symmetries, the minimization procedure can be accomplished more efficiently with only one quarter of the surface, i.e. one face of the pyramid with the appropriate boundary conditions. Fig. \ref{fig:rhombus_accuracy} shows an example of the area convergence as the surface is repeatedly refined. At each refinement, the faces are all divided in four and the new vertices are adjusted. The successive refinements lead to smaller changes in the area, demonstrating convergence.

The coefficients for the $\epsilon$-expansion of the area can be determined through curve fitting. In this example, we consider terms out to order $\epsilon$, yielding three fitting coefficients. Evolving the same surface with five different values $\epsilon=0.03,0.02,0.015,0.01,$ and $0.007$ gives enough data to accurately fix these coefficients.

\begin{figure}[h!]
    \centering
    \includegraphics[width=0.55\textwidth]{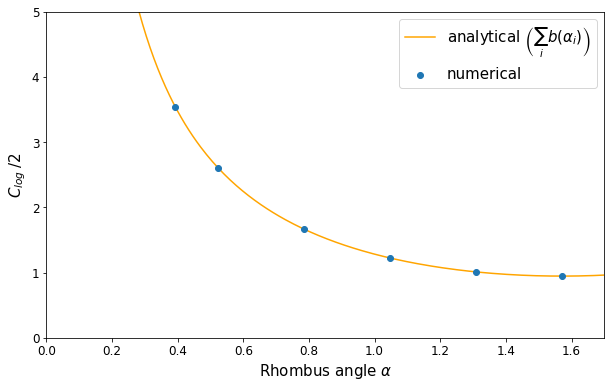}
    \caption{Numerically computed log coefficient for rhombi with side length 1 and vertex angle $\alpha$. These results are an improvement over fig. 5 in \cite{Fonda2015}, since the fitting procedure described above more accurately separates the terms of the $\epsilon$ expansion.}
    \label{fig:rhombus_corners}
\end{figure}

The divergent contribution for corners in AdS$_4$ is analytically known. The $\log$ coefficient for a polygonal region with $N$ vertices is
\begin{equation}
    c_l = 2\sum_{i=1}^N b(\alpha_i),
\end{equation}
where $\alpha_i$ are the angles of the vertices. The function $b$ involves elliptic integrals and is discussed in section 3.2 of \cite{Fonda2015}. As shown in Fig. \ref{fig:rhombus_corners}, the Evolver results are in very close agreement with the analytical result. This represents an improvement over \cite{Fonda2015}, where the series (\ref{rhombus_expansion}) was truncated after the $\log$ term.

\begin{figure}
        \centering
        \includegraphics[width=0.55\textwidth]{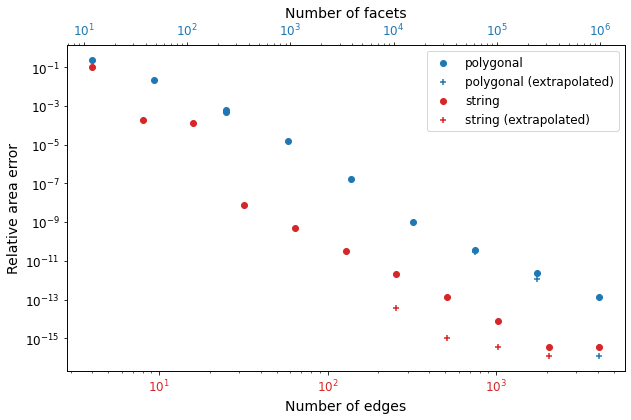}
        \caption{Convergence of minimal surface area with increasing refinement for both the string and polygonal models. In both cases, the cylinder has radius $R=1$ and cutoff $\epsilon=0.01$. The vertical axis shows the relative difference between the numerical and analytical areas. The cross points show values produced using the built in Evolver extrapolation function.}
        \label{fig:cylinder_accuracy}
    \end{figure}

In the remainder of this section we provide additional details for the example in section \ref{sec:inf_cylinder}. The minimal surface for an infinite cylinder can be found numerically using method (1) in \ref{sec:area functional}. However, since the surface can be defined by a function of the radial coordinate only, the string model in \ref{App:SurfaceEvolver} can be used as well. Fig. \ref{fig:cylinder_accuracy} compares the convergence of the areas to the analytical value for both models. By extrapolating the behavior of the last several data points, one can see that both methods converge to the same value to an accuracy within $10^{-15}$ (cross points.) Because the string model reduces the dimensionality of the minimization problem, it can achieve the same accuracy as the polygonal model with less than a hundredth or even a thousandth as many components.
After computing areas for different values of the cutoff, the coefficients can be extracted as before. The log term extracted from fits based on this data matches the analytic result up to an error of order $10^{-5}$. (See (\ref{cyl_results})).

\section{Flows across dimensions in the 5d STU model}\label{App:SugraSolutions}

In this Appendix we provide the details of the supergravity theory that provides holographic flows discussed in section  \ref{sec:evolver_results}. The bosonic sector of 5d $\mathcal N=2$ gauged supergravity known as the STU model is given as
\begin{equation}
\begin{split}
	S&=\fft{1}{16\pi G_{(5)}}\int d^5x\sqrt{|g|}\bigg[R+4\boldsymbol{g}^2\sum_{I=1}^3\fft{1}{X^I}-\fft12\sum_{x=1}^2\partial_\mu\phi^x\partial^\mu\phi^x-\fft14\sum_{I=1}^3(X^I)^{-2}F^I_{\mu\nu}F^I{}^{\mu\nu}\\
	&\kern10em+\fft{1}{4}\varepsilon^{\mu\nu\rho\sigma\lambda}F^1_{\mu\nu}F^2_{\rho\sigma}A^3_\lambda\bigg],
\end{split}\label{N=2:sugra:action:5d}
\end{equation}
where $x \in\{1,2\}$  and $I,J,K\in\{1,2,3\}$. Our convention for the Levi-Civita symbol is 
\begin{equation}
\varepsilon^{\mu\nu\rho\sigma\lambda}=\begin{cases}
	-|g|^{-1/2} & (\text{even permutation})\\
	+|g|^{-1/2} & (\text{odd permutation})
\end{cases}.
\end{equation}
The physical scalars $\phi^x$  parametrize the sections
\begin{equation}
    X^I=e^{\sum_{x=1}^2c^I{}_x\phi^x},
\end{equation}
with the constraint $\sum_{I=1}^3c^I{}_x=0$. Typical values of $c^I{}_x$ are given as
\begin{equation}
    X^1=e^{-\fft{1}{\sqrt6}\phi^1-\fft{1}{\sqrt2}\phi^2},\qquad X^2=e^{-\fft{1}{\sqrt6}\phi^1+\fft{1}{\sqrt2}\phi^2},\qquad X^3=e^{\fft{2}{\sqrt6}\phi^1}.
\end{equation}
%

The Ansatz for the flow solution is 
\begin{align}\label{metric_ansatz}
    ds_5^2&=e^{2f(r)}(-dt^2+dz^2+dr^2) + e^{2g(r)+2h(x,y)}(dx^2+dy^2)\\
    F^I&=-a_I e^{2h(x,y)}dx\wedge dy,\hspace{50 pt}I=1,2,3\nonumber
\end{align}

\begin{equation}
     h(x,y) =
  \begin{cases}
    \frac{1}{2}\log{2\pi}, & \text{for } T^2 \\
    -\log{\frac{1+x^2+y^2}{2}}, & \text{for } S^2
  \end{cases}
\end{equation}
where $x,y\in[0,1)$ for the case of a $T^2$ wrapping \cite{Benini:2013cda}. By defining a new radial variable 

\begin{equation}
    \rho = f + \frac{1}{2\sqrt{6}} \phi_1+ \frac{1}{2\sqrt{2}} \phi_2
\end{equation}
we can rewrite the metric as
\begin{equation} \label{5D sugra metric}
    ds_5^2=e^{2f(\rho)}(-dt^2+dz^2) + \frac{d\rho^2}{D(\rho)^2} + e^{2g(\rho)+2h(x,y)}(dx^2+dy^2)\; ,
\end{equation}
where
\begin{equation}
    D \equiv X^1 + \frac{3a_1}{2}e^{-2g} X_1
\end{equation}
The BPS equations are
\begin{align}
\begin{split}
       0 &= \frac{dg}{d\rho} - \frac1D \Big( \frac{X^1 + X^2 + X^3}3 - e^{-2g} a_I X_I \Big) \\
0 &= \frac{d\phi_1}{d\rho} - \frac{\sqrt6}D \Big( \frac{X^1 + X^2 - 2X^3}3 + e^{-2g} \frac{a_1X_1 + a_2X_2 - 2a_3X_3}2 \Big) \\
0 &= \frac{d\phi_2}{d\rho} - \frac{\sqrt2}D \Big( X^1 - X^2 + 3 e^{-2g} \frac{a_1X_1 - a_2X_2}2 \Big)\\
0 &= a_1 + a_2 + a_3 + \kappa
\end{split}
\end{align}

\bigskip
{\bf Compactification on $\mathbb{T}^2$:}
%
We choose $\{a_1,a_2\}=\{3,3\}$ for the torus compactification. From the last BPS condition, $a_3=-6$. The BPS equations were input into Mathematica and solved using \texttt{WorkingPrecision}$=32$. The definition of $\rho$ was shifted by a constant so that $g(\rho) \rightarrow \rho$ as $\rho\rightarrow\infty$. This convention ensures that the metric (\ref{5D sugra metric}) approaches the standard form for the AdS$_5$ metric in the UV. Numerical data for the functions $f(\rho)$, $g(\rho)$ and $D(\rho)$ was then transferred to the Surface Evolver.


To implement the metric in the Evolver, we first performed coordinate re-definitions
\begin{align}
\begin{split}
    x \rightarrow x_1/\sqrt{2\pi}, \hspace{50 pt}
    y &\rightarrow x_2/\sqrt{2\pi}, \hspace{50 pt}
    z \rightarrow x_3, \\
    \rho &\rightarrow -\log{z}.
\end{split}
\end{align}
This choice gives
\begin{equation}\label{torus metric}
    ds^2=e^{2f(\rho)}(-dt^2+dx_3^2) + \frac{dz^2}{z^2 D(\rho)^2} + e^{2g(\rho)}(dx_1^2+dx_2^2)\;,
\end{equation}
where $x_1,x_2 \in \left[-\sqrt{\pi/2},\sqrt{\pi/2}\right)$ and $\rho$ is considered a function of $z$. The data for functions $f$, $g$ and $D$ are still indexed by $\rho$ in the code since the interpolation can by done more accurately than if it were given by equally spaced values in $z$.

For an entangling region that symmetrically wraps the direction $x_1$, the minimal hypersurface is given by $z=z(x_2, x_3)$. The volume of the hypersurface at constant timeslice is
\begin{equation}\label{torus area integral}
    \sqrt{2\pi}\bigintss_{\cal A} {e^{g}\sqrt{e^{2f+2g}+e^{2f}\frac{1}{z^2 D^2}\left(\frac{\partial z}{\partial x_2}\right)^2 + e^{2g}\frac{1}{z^2 D^2}\left(\frac{\partial z}{\partial x_3}\right)^2}dx_2 dx_3}
\end{equation}
where the integration region is the cross section of the entangling region in the $x_2,x_3$ plane. The area minimization problem can then be solved with a two-dimensional surface in ($x_2,x_3,z$) space with the integral above replacing the usual area.

\bigskip
{\bf Compactification on $S^2$:}
\begin{figure}
    \centering
    \includegraphics[width=0.55\textwidth]{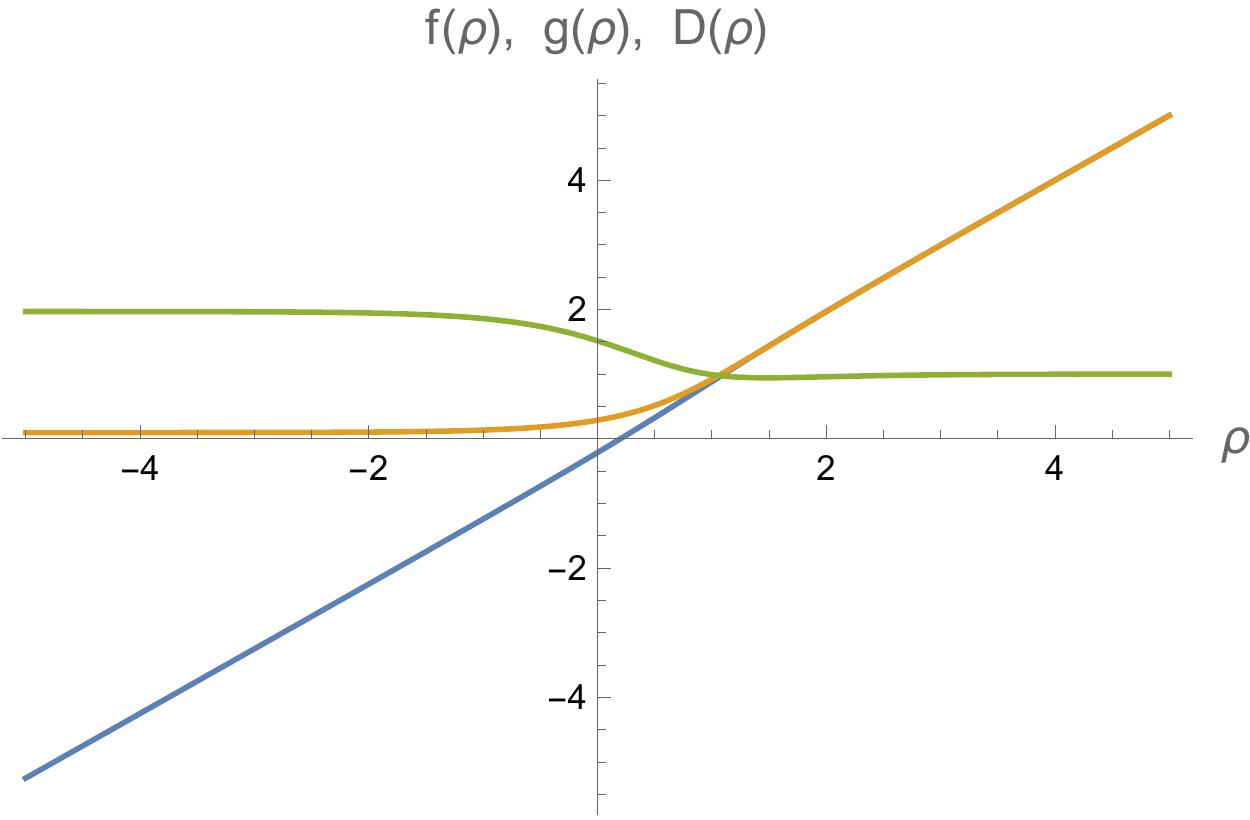}
    \caption{Flow functions for the $\mathbb{S}^2$ compactification. The curves for $f$, $g$ and $D$ are blue, orange, and green respectively.}
    \label{fig:sphere_fgd}
\end{figure}
For compactification on a sphere we take $\{a_1,a_2,a_3\}=\{2,2,-5\}$. The setup is similar to the torus case, but with the metric
\begin{equation}\label{sphere metric}
    ds^2=e^{2f(\rho)}(-dt^2+dx^2) + \frac{dz^2}{z^2 D(\rho)^2} + e^{2g(\rho)}(d\theta^2+\sin^2\theta d\phi^2)\;,
\end{equation}
where $\theta\in[0,\pi]$ and $\phi\in[0,2\pi)$. For an entangling region symmetric with respect to $\phi$, the area integral for the minimal surface must also be independent of $\phi$. The area is
\begin{equation}\label{sphere area integral}
    2\pi\bigintss_{\cal A} {e^{g}\sin(\theta)\sqrt{e^{2f+2g}+e^{2f}\frac{1}{z^2 D^2}\left(\frac{\partial z}{\partial\theta}\right)^2 + e^{2g}\frac{1}{z^2 D^2}\left(\frac{\partial z}{\partial x}\right)^2}dx d\theta}
\end{equation}
As before, we can now consider the minimization problem with a surface in $(x,\theta,z)$ space.

\bibliographystyle{JHEP}
\bibliography{EE-refs.bib}

\providecommand{\href}[2]{#2}\begingroup\raggedright\begin{thebibliography}{10}

\bibitem{Zamolodchikov:1986gt}
A.~B. Zamolodchikov, \emph{{Irreversibility of the Flux of the Renormalization
  Group in a 2D Field Theory}}, {\emph{JETP Lett.} {\bfseries 43} (1986)
  730--732}.

\bibitem{Cardy:1988cwa}
J.~L. Cardy, \emph{{Is There a c Theorem in Four-Dimensions?}},
  \href{https://doi.org/10.1016/0370-2693(88)90054-8}{\emph{Phys. Lett. B}
  {\bfseries 215} (1988) 749--752}.

\bibitem{Komargodski:2011vj}
Z.~Komargodski and A.~Schwimmer, \emph{{On Renormalization Group Flows in Four
  Dimensions}}, \href{https://doi.org/10.1007/JHEP12(2011)099}{\emph{JHEP}
  {\bfseries 12} (2011) 099},
  [\href{https://arxiv.org/abs/1107.3987}{{\ttfamily 1107.3987}}].

\bibitem{Jafferis:2010un}
D.~L. Jafferis, \emph{{The Exact Superconformal R-Symmetry Extremizes Z}},
  \href{https://doi.org/10.1007/JHEP05(2012)159}{\emph{JHEP} {\bfseries 05}
  (2012) 159}, [\href{https://arxiv.org/abs/1012.3210}{{\ttfamily 1012.3210}}].

\bibitem{Klebanov:2011gs}
I.~R. Klebanov, S.~S. Pufu and B.~R. Safdi, \emph{{F-Theorem without
  Supersymmetry}}, \href{https://doi.org/10.1007/JHEP10(2011)038}{\emph{JHEP}
  {\bfseries 10} (2011) 038},
  [\href{https://arxiv.org/abs/1105.4598}{{\ttfamily 1105.4598}}].

\bibitem{Jafferis:2011zi}
D.~L. Jafferis, I.~R. Klebanov, S.~S. Pufu and B.~R. Safdi, \emph{{Towards the
  F-Theorem: N=2 Field Theories on the Three-Sphere}},
  \href{https://doi.org/10.1007/JHEP06(2011)102}{\emph{JHEP} {\bfseries 06}
  (2011) 102}, [\href{https://arxiv.org/abs/1103.1181}{{\ttfamily 1103.1181}}].

\bibitem{Casini:2012ei}
H.~Casini and M.~Huerta, \emph{{On the RG running of the entanglement entropy
  of a circle}}, \href{https://doi.org/10.1103/PhysRevD.85.125016}{\emph{Phys.
  Rev. D} {\bfseries 85} (2012) 125016},
  [\href{https://arxiv.org/abs/1202.5650}{{\ttfamily 1202.5650}}].

\bibitem{Casini:2006es}
H.~Casini and M.~Huerta, \emph{{A c-theorem for the entanglement entropy}},
  \href{https://doi.org/10.1088/1751-8113/40/25/S57}{\emph{J. Phys. A}
  {\bfseries 40} (2007) 7031--7036},
  [\href{https://arxiv.org/abs/cond-mat/0610375}{{\ttfamily
  cond-mat/0610375}}].

\bibitem{Casini:2017roe}
H.~Casini, E.~Teste and G.~Torroba, \emph{{Modular Hamiltonians on the null
  plane and the Markov property of the vacuum state}},
  \href{https://doi.org/10.1088/1751-8121/aa7eaa}{\emph{J. Phys. A} {\bfseries
  50} (2017) 364001}, [\href{https://arxiv.org/abs/1703.10656}{{\ttfamily
  1703.10656}}].

\bibitem{Casini:2017vbe}
H.~Casini, E.~Test\'e and G.~Torroba, \emph{{Markov Property of the Conformal
  Field Theory Vacuum and the a Theorem}},
  \href{https://doi.org/10.1103/PhysRevLett.118.261602}{\emph{Phys. Rev. Lett.}
  {\bfseries 118} (2017) 261602},
  [\href{https://arxiv.org/abs/1704.01870}{{\ttfamily 1704.01870}}].

\bibitem{Maldacena:2000mw}
J.~M. Maldacena and C.~Nunez, \emph{{Supergravity description of field theories
  on curved manifolds and a no go theorem}},
  \href{https://doi.org/10.1142/S0217751X01003937}{\emph{Int. J. Mod. Phys. A}
  {\bfseries 16} (2001) 822--855},
  [\href{https://arxiv.org/abs/hep-th/0007018}{{\ttfamily hep-th/0007018}}].

\bibitem{Acharya:2000mu}
B.~S. Acharya, J.~P. Gauntlett and N.~Kim, \emph{{Five-branes wrapped on
  associative three cycles}},
  \href{https://doi.org/10.1103/PhysRevD.63.106003}{\emph{Phys. Rev. D}
  {\bfseries 63} (2001) 106003},
  [\href{https://arxiv.org/abs/hep-th/0011190}{{\ttfamily hep-th/0011190}}].

\bibitem{Gauntlett:2000ng}
J.~P. Gauntlett, N.~Kim and D.~Waldram, \emph{{M Five-branes wrapped on
  supersymmetric cycles}},
  \href{https://doi.org/10.1103/PhysRevD.63.126001}{\emph{Phys. Rev. D}
  {\bfseries 63} (2001) 126001},
  [\href{https://arxiv.org/abs/hep-th/0012195}{{\ttfamily hep-th/0012195}}].

\bibitem{Gauntlett:2001qs}
J.~P. Gauntlett, N.~Kim, S.~Pakis and D.~Waldram, \emph{{Membranes wrapped on
  holomorphic curves}},
  \href{https://doi.org/10.1103/PhysRevD.65.026003}{\emph{Phys. Rev. D}
  {\bfseries 65} (2002) 026003},
  [\href{https://arxiv.org/abs/hep-th/0105250}{{\ttfamily hep-th/0105250}}].

\bibitem{Gauntlett:2001jj}
J.~P. Gauntlett and N.~Kim, \emph{{M five-branes wrapped on supersymmetric
  cycles. 2.}}, \href{https://doi.org/10.1103/PhysRevD.65.086003}{\emph{Phys.
  Rev. D} {\bfseries 65} (2002) 086003},
  [\href{https://arxiv.org/abs/hep-th/0109039}{{\ttfamily hep-th/0109039}}].

\bibitem{Benini:2013cda}
F.~Benini and N.~Bobev, \emph{{Two-dimensional SCFTs from wrapped branes and
  c-extremization}}, \href{https://doi.org/10.1007/JHEP06(2013)005}{\emph{JHEP}
  {\bfseries 06} (2013) 005},
  [\href{https://arxiv.org/abs/1302.4451}{{\ttfamily 1302.4451}}].

\bibitem{Benini:2015bwz}
F.~Benini, N.~Bobev and P.~M. Crichigno, \emph{{Two-dimensional SCFTs from
  D3-branes}}, \href{https://doi.org/10.1007/JHEP07(2016)020}{\emph{JHEP}
  {\bfseries 07} (2016) 020},
  [\href{https://arxiv.org/abs/1511.09462}{{\ttfamily 1511.09462}}].

\bibitem{Bobev:2017uzs}
N.~Bobev and P.~M. Crichigno, \emph{{Universal RG Flows Across Dimensions and
  Holography}}, \href{https://doi.org/10.1007/JHEP12(2017)065}{\emph{JHEP}
  {\bfseries 12} (2017) 065},
  [\href{https://arxiv.org/abs/1708.05052}{{\ttfamily 1708.05052}}].

\bibitem{Bea:2015fja}
Y.~Bea, J.~D. Edelstein, G.~Itsios, K.~S. Kooner, C.~Nunez, D.~Schofield
  et~al., \emph{{Compactifications of the Klebanov-Witten CFT and new AdS$_{3}$
  backgrounds}}, \href{https://doi.org/10.1007/JHEP05(2015)062}{\emph{JHEP}
  {\bfseries 05} (2015) 062},
  [\href{https://arxiv.org/abs/1503.07527}{{\ttfamily 1503.07527}}].

\bibitem{GonzalezLezcano:2022mcd}
A.~Gonz\'alez~Lezcano, J.~Hong, J.~T. Liu, L.~A. Pando~Zayas and C.~F.
  Uhlemann, \emph{{c-functions in flows across dimensions}},
  \href{https://doi.org/10.1007/JHEP10(2022)083}{\emph{JHEP} {\bfseries 10}
  (2022) 083}, [\href{https://arxiv.org/abs/2207.09360}{{\ttfamily
  2207.09360}}].

\bibitem{Brakke}
K.~A. Brakke, \emph{http://facstaff.susqu.edu/brakke/evolver/evolver.html}, .

\bibitem{Brakke1992}
K.~A. Brakke, \emph{The surface evolver},
  \href{https://doi.org/10.1080/10586458.1992.10504253}{\emph{Experimental
  Mathematics} {\bfseries 1} (1992) 141--165},
  [\href{https://arxiv.org/abs/https://doi.org/10.1080/10586458.1992.10504253}{{\ttfamily
  https://doi.org/10.1080/10586458.1992.10504253}}].

\bibitem{Srednicki1993}
M.~Srednicki, \emph{Entropy and area},
  \href{https://doi.org/10.1103/physrevlett.71.666}{\emph{Physical Review
  Letters} {\bfseries 71} (1993) 666--669},
  [\href{https://arxiv.org/abs/9303048}{{\ttfamily 9303048}}].

\bibitem{Casini2009}
H.~Casini and M.~Huerta, \emph{Entanglement entropy in free quantum field
  theory}, \href{https://doi.org/10.1088/1751-8113/42/50/504007}{\emph{Journal
  of Physics A: Mathematical and Theoretical} {\bfseries 42} (2009) 504007},
  [\href{https://arxiv.org/abs/0905.2562}{{\ttfamily 0905.2562}}].

\bibitem{Witten:2018zxz}
E.~Witten, \emph{{APS Medal for Exceptional Achievement in Research: Invited
  article on entanglement properties of quantum field theory}},
  \href{https://doi.org/10.1103/RevModPhys.90.045003}{\emph{Rev. Mod. Phys.}
  {\bfseries 90} (2018) 045003},
  [\href{https://arxiv.org/abs/1803.04993}{{\ttfamily 1803.04993}}].

\bibitem{Myers2012}
R.~C. Myers and A.~Singh, \emph{Entanglement entropy for singular surfaces},
  \href{https://doi.org/10.1007/jhep09(2012)013}{\emph{Journal of High Energy
  Physics} {\bfseries 2012} (2012) },
  [\href{https://arxiv.org/abs/1206.5225}{{\ttfamily 1206.5225}}].

\bibitem{Ryu:2006bv}
S.~Ryu and T.~Takayanagi, \emph{{Holographic derivation of entanglement entropy
  from AdS/CFT}},
  \href{https://doi.org/10.1103/PhysRevLett.96.181602}{\emph{Phys. Rev. Lett.}
  {\bfseries 96} (2006) 181602},
  [\href{https://arxiv.org/abs/hep-th/0603001}{{\ttfamily hep-th/0603001}}].

\bibitem{Ryu:2006ef}
S.~Ryu and T.~Takayanagi, \emph{{Aspects of holographic entanglement entropy}},
  {\emph{JHEP} {\bfseries 08} (2006) 045},
  [\href{https://arxiv.org/abs/hep-th/0605073}{{\ttfamily hep-th/0605073}}].

\bibitem{Nishioka:2009un}
T.~Nishioka, S.~Ryu and T.~Takayanagi, \emph{{Holographic Entanglement Entropy:
  An Overview}}, \href{https://doi.org/10.1088/1751-8113/42/50/504008}{\emph{J.
  Phys. A} {\bfseries 42} (2009) 504008},
  [\href{https://arxiv.org/abs/0905.0932}{{\ttfamily 0905.0932}}].

\bibitem{10.2307/1989472}
J.~Douglas, \emph{Solution of the problem of plateau}, {\emph{Transactions of
  the American Mathematical Society} {\bfseries 33} (1931) 263--321}.

\bibitem{10.2307/1968237}
T.~Rado, \emph{On plateau's problem}, {\emph{Annals of Mathematics} {\bfseries
  31} (1930) 457--469}.

\bibitem{Fonda2015}
P.~Fonda, L.~Giomi, A.~Salvio and E.~Tonni, \emph{On shape dependence of
  holographic mutual information in {AdS}4},
  \href{https://doi.org/10.1007/jhep02(2015)005}{\emph{Journal of High Energy
  Physics} {\bfseries 2015} (2015) },
  [\href{https://arxiv.org/abs/1411.3608}{{\ttfamily 1411.3608}}].

\bibitem{Seminara2017}
D.~Seminara, J.~Sisti and E.~Tonni, \emph{Corner contributions to holographic
  entanglement entropy in {AdS}4/{BCFT}3},
  \href{https://doi.org/10.1007/jhep11(2017)076}{\emph{Journal of High Energy
  Physics} {\bfseries 2017} (2017) },
  [\href{https://arxiv.org/abs/1708.05080}{{\ttfamily 1708.05080}}].

\bibitem{Solodukhin:2008dh}
S.~N. Solodukhin, \emph{{Entanglement entropy, conformal invariance and
  extrinsic geometry}},
  \href{https://doi.org/10.1016/j.physletb.2008.05.071}{\emph{Phys. Lett. B}
  {\bfseries 665} (2008) 305--309},
  [\href{https://arxiv.org/abs/0802.3117}{{\ttfamily 0802.3117}}].

\bibitem{Liu:2012eea}
H.~Liu and M.~Mezei, \emph{{A Refinement of entanglement entropy and the number
  of degrees of freedom}},
  \href{https://doi.org/10.1007/JHEP04(2013)162}{\emph{JHEP} {\bfseries 04}
  (2013) 162}, [\href{https://arxiv.org/abs/1202.2070}{{\ttfamily 1202.2070}}].

\bibitem{Uhlemann:2021itz}
C.~F. Uhlemann, \emph{{Information transfer with a twist}},
  \href{https://doi.org/10.1007/JHEP01(2022)126}{\emph{JHEP} {\bfseries 01}
  (2022) 126}, [\href{https://arxiv.org/abs/2111.11443}{{\ttfamily
  2111.11443}}].

\bibitem{Klemm:2000nj}
D.~Klemm and W.~A. Sabra, \emph{{Supersymmetry of black strings in D = 5 gauged
  supergravities}},
  \href{https://doi.org/10.1103/PhysRevD.62.024003}{\emph{Phys. Rev. D}
  {\bfseries 62} (2000) 024003},
  [\href{https://arxiv.org/abs/hep-th/0001131}{{\ttfamily hep-th/0001131}}].

\bibitem{Macpherson:2014eza}
N.~T. Macpherson, C.~N\'u\~nez, L.~A. Pando~Zayas, V.~G.~J. Rodgers and C.~A.
  Whiting, \emph{{Type IIB supergravity solutions with AdS$_{5}$ from Abelian
  and non-Abelian T dualities}},
  \href{https://doi.org/10.1007/JHEP02(2015)040}{\emph{JHEP} {\bfseries 02}
  (2015) 040}, [\href{https://arxiv.org/abs/1410.2650}{{\ttfamily 1410.2650}}].

\bibitem{Chu:2019uoh}
C.-S. Chu and D.~Giataganas, \emph{{$c$-Theorem for Anisotropic RG Flows from
  Holographic Entanglement Entropy}},
  \href{https://doi.org/10.1103/PhysRevD.101.046007}{\emph{Phys. Rev. D}
  {\bfseries 101} (2020) 046007},
  [\href{https://arxiv.org/abs/1906.09620}{{\ttfamily 1906.09620}}].

\bibitem{Legramandi:2021aqv}
A.~Legramandi and C.~Nunez, \emph{{Holographic description of SCFT$_{5}$
  compactifications}},
  \href{https://doi.org/10.1007/JHEP02(2022)010}{\emph{JHEP} {\bfseries 02}
  (2022) 010}, [\href{https://arxiv.org/abs/2109.11554}{{\ttfamily
  2109.11554}}].

\bibitem{Merrikin:2022yho}
P.~Merrikin, C.~Nunez and R.~Stuardo, \emph{{Compactification of 6d ${\cal
  N}=(1,0)$ quivers, 4d SCFTs and their holographic dual Massive IIA
  backgrounds}},  \href{https://arxiv.org/abs/2210.02458}{{\ttfamily
  2210.02458}}.

\end{thebibliography}\endgroup

\end{document}